\documentclass{article}

\usepackage{arxiv}

\usepackage[utf8]{inputenc} 
\usepackage[T1]{fontenc}    
\usepackage{hyperref}       
\usepackage{url}            
\usepackage{booktabs}       
\usepackage{amsfonts}       
\usepackage{nicefrac}       
\usepackage{microtype}      
\usepackage{lipsum}

\usepackage{enumerate}
\usepackage{amsmath}
\usepackage{amssymb}
\usepackage{graphicx}

\usepackage{algorithm}    
\usepackage{algpseudocode}  
\usepackage{setspace}       

\usepackage{etoolbox}
\usepackage{lipsum}

\makeatletter
\newcommand*{\algrule}[1][\algorithmicindent]{%
  \hspace*{0.2em}
  \vrule 
  \hspace*{\dimexpr#1-.2em-.4pt}%
}

\newcommand{\StatePar}[1]{%
  \State\parbox[t]{\dimexpr\linewidth-\ALG@thistlm}{\strut #1\strut}%
}
\renewcommand{\ALG@beginalgorithmic}{\offinterlineskip}

\newcount\ALG@printindent@tempcnta
\def\ALG@printindent{%
  \ifnum \theALG@nested > 0
    \ifx\ALG@text\ALG@x@notext
    \else
      \unskip
      \ALG@printindent@tempcnta=1
      \loop
        \algrule[\csname ALG@ind@\the\ALG@printindent@tempcnta\endcsname]%
        \advance \ALG@printindent@tempcnta 1
        \ifnum \ALG@printindent@tempcnta<\numexpr\theALG@nested+1\relax
      \repeat
        \fi
    \fi
}
\patchcmd{\ALG@doentity}{\noindent\hskip\ALG@tlm}{\ALG@printindent}{}{\errmessage{failed to patch}}
\makeatother

\algrenewcommand\algorithmicend{\strut\textbf{end}}
\algrenewcommand\algorithmicdo{\strut\textbf{do}}
\algrenewcommand\algorithmicwhile{\strut\textbf{while}}
\algrenewcommand\algorithmicfor{\strut\textbf{for}}
\algrenewcommand\algorithmicforall{\strut\textbf{for all}}
\algrenewcommand\algorithmicloop{\strut\textbf{loop}}
\algrenewcommand\algorithmicrepeat{\strut\textbf{repeat}}
\algrenewcommand\algorithmicuntil{\strut\textbf{until}}
\algrenewcommand\algorithmicprocedure{\strut\textbf{procedure}}
\algrenewcommand\algorithmicfunction{\strut\textbf{function}}
\algrenewcommand\algorithmicif{\strut\textbf{if}}
\algrenewcommand\algorithmicthen{\strut\textbf{then}}
\algrenewcommand\algorithmicelse{\strut\textbf{else}}

\algrenewcommand\algorithmicrequire{\strut\textbf{Input:}}
\algrenewcommand\algorithmicensure{\strut\textbf{Output:}}

\let\oldState\State
\renewcommand{\State}{\oldState\strut}

\usepackage{booktabs}
\usepackage{multirow}

\begin{document}


\author{Angel Gim\'{e}nez$^a$\footnote{Corresponding author}, Miguel A. Murcia, Jos\'{e} M. Amig\'{o}$^a$, Oscar Mart\'{\i}nez-Bonastre$^a$ and Jos{\'e} Valero$^a$\\
 $^a$Centro de Investigaci\'{o}n Operativa \\
Universidad Miguel Hern\'{a}ndez de Elche, \\ 
Avda. de la Universidad s/n, 03202 Elche, Spain \\
\texttt{a.gimenez@umh.es, miguel.murcia02@graduado.umh.es, jm.amigo@umh.es,} \\
\texttt{oscar.martinez@umh.es, jvalero@umh.es}

}

\title{New RED-type TCP-AQM algorithms based on beta distribution drop functions}

\maketitle

\begin{abstract}
In recent years, \textit{Active Queue Management} (AQM) mechanisms to improve the performance of TCP/IP networks have acquired a relevant role. In this paper we present a simple and robust RED-type algorithm together with a couple of dynamical variants  with the ability to adapt to the specific characteristics of different network environments, as well as to the user's needs. We first present a basic version called  \textit{Beta RED} (BetaRED), where the user is free to adjust the parameters according to the network conditions. The aim is to make the parameter setting easy and intuitive so that a good performance is obtained over a wide range of parameters. Secondly, BetaRED is used as a framework to design two dynamic algorithms, which we will call \textit{Adaptive Beta RED} (ABetaRED) and \textit{Dynamic Beta RED} (DBetaRED). In those new algorithms certain parameters are dynamically adjusted so that the queue length remains stable around a predetermined reference value and according to changing network traffic conditions. Finally, we present a battery of simulations using the \textit{Network Simulator 3} (ns-3) software with a two-fold objective: to guide the user on how to adjust the parameters of the BetaRED mechanism, and to show a performance comparison of ABetaRED and DBetaRED with other representative algorithms that pursue a similar objective.
\end{abstract}


\section{Introduction}
\label{sec-introduction}

Over the last few decades, the Internet has become increasingly faster by many orders of magnitude, but at the same time the number of users has increased, along with a huge deployment of Internet-based applications, and with a multitude of bandwidth and latency requirements. This growth in data flow has resulted in increased network traffic congestion, which if it is not properly managed could lead to a considerable decrease in performance, a critical factor for any Internet service. A very topical problem, for example, is the massive use of \textit{Voice over Internet Protocol} (VoIP) applications, online games, financial exchanges, etc., which cause buffers to fill up completely, resulting in the phenomenon known as bufferbloat \cite{gettys_bufferbloat_2011}, characterized by high latency and massive discarding of packets over long periods of time, leading to poor quality of service for the end user.

Congestion control is a critical part of the Transmission Control Protocol (TCP), which directly influences transport performance. TCP is the dominant transport layer protocol on the Internet. In general, there are two congestion control mechanisms: (1) End-to-end congestion controls; these are approaches adopted by TCP and are achieved mainly at the transport layer. (2) Network-assisted congestion controls; these are controls adopted by routers. This mechanism uses router queue size and/or delay to monitor the congestion state of the network. To meet the demands of Internet users and applications, research both in academic and industrial environments has focused on improving these two control mechanisms to achieve high performance and avoid data traffic collapse. This paper is concerned with the second approach, and aims to design a new congestion control mechanism to be implemented at the router. For those readers interested in the first approach, the surveys \cite{al-saadi_survey_2019, polese_survey_2019} collect the main transport protocols that have been proposed in recent years.

The main objective of congestion control mechanisms is to keep the network operating fairly close to its nominal capacity, even when faced with extreme overload. This objective is achieved by acting according to two fundamental premises. The first is to prevent network congestion before it occurs and to dissolve congestion if it cannot be avoided. The second is to provide a fairness service to different connections, along with support for various Internet application domains with varying quality of service (QoS) requirements. Designing good congestion control mechanisms is extremely difficult; the specific characteristics of the connection and each individual link can significantly affect performance, often in unpredictable ways.

\textit{Active Queue Management} (AQM) algorithms are algorithms implemented in routers that act on the buffer queue to control its length so that efficient congestion control of the system is achieved. Such management also allows TCP to do its job of sharing links properly, without which it cannot function as intended. The primary goal of an AQM is to prevent congestion before it occurs and, if it has already occurred, to try to control it. AQM mechanisms act on the length of queues in buffers to achieve lower delay, and try to absorb short-term variations (e.g., bursts), thus playing  a crucial role in congestion control on the Internet. Numerous AQM schemes have been proposed in recent years to properly manage queues to avoid undesirable effects such as bufferbloat, link underutilization, large variations in queuing delays, etc. The truth is that there is no one algorithm that has dominated over the others, since network environments are very varied, and network traffic very changeable, so each algorithm has advantages and disadvantages according to those environments.

The \textit{Random Early Detection} (RED) algorithm, proposed by Floyd and Jacobson \cite{floyd_random_1993}, works by detecting incipient congestion and notifying the TCP transmission control protocol of the congestion by randomly discarding packets to avoid filling the router. The RED algorithm avoids some of the problems of using the simple \textit{Drop Tail} \cite{shorten_modelling_2007} (such as blocking, full queues and global synchronization), whose mechanism consists of dropping all incoming packets when the buffer queue is full. In fact, RED has been the most studied AQM to date, and has been the basis for the development of new AQM systems. The reason for this is not only because RED was the first to be developed in the Internet community, but also because of the numerous drawbacks involved in using this algorithm, some of which have not yet been fully resolved. One of the main problems is the difficulty and uncertainty in adjusting its parameters for adequate performance, due to the high sensitivity of the RED parameters to traffic load. Thus, a poor choice of RED parameters can lead to other deficiencies such as forced drops, or link underutilization. Moreover, even if the RED parameters are properly tuned, they are very sensitive to network conditions and can also cause other more complex nonlinear effects such as bifurcation and chaos.  Summarizing, it can happen that a set of parameters works perfectly for a certain network setting but not when those parameters are slightly changed. This is obviously not desirable, as Internet traffic conditions change rapidly. The study of stability and bifurcations of Internet congestion control models involving delay is a central issue that has been intensively studied in recent decades (see \cite{chen_controlling_2004,liu_controlling_2010,xiao_bifurcation_2013,zhang_stability_2015,pei_dynamics_2019,duran_bifurcation_2019, m_amigo_generalized_2020,amigo_modeling_2021,pei_periodic_2021}).

The paper is organized as follows. In Section \ref{sec-related-works} the current AQM problem is presented, along with the description of the most recent algorithms that have similar objectives to BetaRED. This section also briefly reviews the AQM schemes with which we will compare the proposed new algorithms. Section \ref{sec-scenarios-metrics} is devoted to describing the simulation scenarios and the metrics we will use to perform the comparison between the AQM schemes. In the next two sections, new proposed algorithms are introduced and compared by numerical simulations with ns-3, namely, the main BetaRED algorithm (Section \ref{sec-beta-red}), and two new dynamic algorithms ABetaRED and DBetaRED based on the previous one (Section \ref{sec-dynamic-beta-red}). Finally, in Section \ref{sec-conclusions}, some final conclusions are presented.

For the convenience of the reader, in Table \ref{table-notation-parameters} we list symbols of the most important parameters used in this paper.

\begin{table}[htbp]
\renewcommand{\arraystretch}{1.4}
\centering
\begin{tabular}{l|p{3cm}|p{1.5cm}|p{6cm}|}
	\cline{2-4}
                                                          & AQM  algorithm                                                 & Name                  & Description                                                                                                                   \\ \hline
\multicolumn{1}{|l|}{\multirow{9}{2cm}{Tunable parameters}} & \multirow{3}{3cm}{ABetaRED, DBetaRED, ARED, CoDel, PIE} & $T_{\mathrm{target}}$ & Target delay                                                                                                                  \\ \cline{3-4} 
\multicolumn{1}{|l|}{}                                    &                                                       & $T_{\mathrm{update}}$ & Update interval time                                                                                                          \\ \cline{3-4} 
\multicolumn{1}{|l|}{}                                    &                                                       & Alpha, Beta           & Control parameters with different objectives according to the AQM.                                                            \\ \cline{2-4} 
\multicolumn{1}{|l|}{}                                    & BetaRED, ABetaRED, DBetaRED                           & $\theta$              & Scale factor determining the standard deviation of the drop probability function                                                                               \\ \cline{3-4} 
\multicolumn{1}{|l|}{}                                    &                                     & $w$                   & Averaging weight                                                                                                            \\ \cline{2-4} 
\multicolumn{1}{|l|}{}                                    & \multirow{4}{*}{BetaRED}                              & $q_{\mathrm{target}}$ & Target queue length                                                                                                                  \\ \cline{3-4} 
\multicolumn{1}{|l|}{}                                    &                                                       & $q_{\min}$            & Lower threshold                                                                                                               \\ \cline{3-4} 
\multicolumn{1}{|l|}{}                                    &                                                       & $q_{\max}$            & Upper threshold                                                                                                               \\ \cline{3-4} 
\multicolumn{1}{|l|}{}                                    &                                                       & $p_{\max}$            & Maximum packet drop probability                                                                                               \\ \hline
\multicolumn{1}{|l|}{\multirow{4}{2cm}{System parameters}}  & \multirow{4}{*}{All}                                  & $B$                   & Buffer size (maximum number of packets that the buﬀer of Router 1 can store)                                                  \\ \cline{3-4} 
\multicolumn{1}{|l|}{}                                    &                                                       & $C$                   & Capacity of the channel (the maximum amount of error-free information that can be transmitted over the channel per unit time) \\ \cline{3-4} 
\multicolumn{1}{|l|}{}                                    &                                                       & $N$                   & Number of flows in the dumbbell topology                                                                                      \\ \cline{3-4} 
\multicolumn{1}{|l|}{}                                    &                                                       & $M$                   & Packet size                                                                                                                   \\ \hline
\multicolumn{1}{|l|}{\multirow{3}{*}{Variables}}          & \multirow{3}{*}{All}                                  & $p$                   & Drop probability                                                                                                              \\ \cline{3-4} 
\multicolumn{1}{|l|}{}                                    &                                                       & $q_{\mathrm{cur}}$    & Current queue length at Router 1                                                                                              \\ \cline{3-4}
\multicolumn{1}{|l|}{}                                    &                                                       & $q_{\mathrm{avg}}$    & Average queue length at Router 1                                                                                              \\ \hline

\end{tabular}
\caption{Notation for the most important parameters and variables used throughout the paper.}
\label{table-notation-parameters}
\end{table}

\section{Description of the problem and related works} \label{sec-related-works}

Numerous AQM algorithms with different approaches have been proposed over the past two decades to handle the queuing delay problem. However, the conditions of a network and the specific user needs may vary from one scenario to another, so there is no algorithm that can satisfy all demands at the same time. In addition, the optimal configuration of the parameters involved in this type of algorithm is complicated. This is why no single AQM has had a predominant deployment over the others, and the AQM that obtains the best performance in the given scenario is selected. This fact is what has most negatively affected the widespread application of the RED mechanism. There are a large number of publications in the literature aimed at overcoming these difficulties by variations on the RED algorithm. A good summary can be found in \cite{adams_active_2013}, which in turn contains a large collection of citations of work based on RED technology. However, the problem is still topical and we can find numerous recent works addressing the issue.

Drop Tail is the simplest algorithm that can be designed: the buffer accepts packets until it is completely full, and when this happens, it discards the last packets received. The main problem with this mechanism is its performance when combined with TCP, which reduces the sending rate when it receives packet loss notifications, and this occurs only after the buffer is completely full. This, in turn, causes TCP to generate a flow with intermittent bursts, which, with full buffers, causes packet loss to increase in each \textit{Round Trip Time} (RTT) cycle, eventually ending up in a  degraded circular dynamics with high latency and low throughput. The problem is further aggravated when the number of flows connected in the same link increases, each regulated by its own congestion window. It has been found that in this situation, the slow start of each of the flows tends to synchronize, causing the phenomenon known as global synchronization \cite{floyd_random_1993}, which is characterized by very low overall network throughput.

Active queue management attempts to provide a solution to problems that appear in Drop Tail  (see \cite{brandauer_comparison_2001,shorten_modelling_2007}) by notifying congestion initiation signals before the buffer fills up. Notification can be done via packet drop or with the \textit{Explicit Congestion Notification} (ECN) flag. The large amount of research on the design and study of AQM algorithms has resulted in significant improvement in performance metrics: latency, link utilization, throughput, jitter, etc., reducing burst losses and solving the global synchronization problem.

In the RED scheme, the probability function $p$ of packet dropping at an instant is dependent on the average size of the queue length $q_{\mathrm{avg}}$ and its expression is given by 
\begin{equation}  \label{p-ave}
	p(q_{\mathrm{avg}})= 
	\begin{cases}
	0 & \text{if }q_{\mathrm{avg}}<q_{\min }, \\ 
	1 & \text{if }q_{\mathrm{avg}}>q_{\max }, \\ 
	p_{\max }\cdot z(q_{\mathrm{avg}}) & \text{otherwise.}
	\end{cases} \qquad\text{where}\quad
	z(q_{\mathrm{avg}})=\frac{q_{\mathrm{avg}}-q_{\min}}{q_{\max}-q_{\min}}
\end{equation}
The values $q_{\min}$ and $q_{\max}$ are the lower and upper thresholds for the average queue length $q_{\mathrm{avg}}$ where below $q_{\min}$ all packets are accepted, and above $q_{\max}$ all packets are rejected. The value $p_{\max}$ is the maximum value for the packet drop probability function when the average queue length is between $q_{\min}$ and $q_{\max}$, which is reached at the point $q_{\mathrm{avg}}=q_{\max }$. The average queue length is updated upon packet arrival according to the \textit{Exponential Weighted Moving Average} (EWMA),
\begin{equation}\label{q-new}
	q_{\mathrm{avg}}^{\mathrm{new}}=(1-w)\cdot q_{\mathrm{avg}}^{\mathrm{old}}+w\cdot q_{\mathrm{cur}}  
\end{equation} 
between the previous average queue length $q_{\mathrm{avg}}^{\mathrm{old}}$
and the current queue length $q_{\mathrm{cur}}$, where $0<w<1$ is the \textit{averaging weight}. The higher $w$, the faster the RED mechanism reacts to
the actual buffer occupancy.

Although RED was able to eliminate some shortcomings of Drop Tail such as blocking, full queues and global synchronization, other shortcomings remained. One of the main problems of RED is that its parameters must be adjusted according to different Internet traffic load states.
Experiments and numerical simulations have been the main tool used to adjust the parameters of many of the algorithms studied. In the literature we find many conclusions based solely on simulation results, but they lack a theoretical foundation that guarantees the results. The combination of end-to-end TCP congestion control and active RED queue management can be modeled as a discrete-time dynamic system and this system exhibits a variety of irregular behaviors, such as bifurcation and chaos. Recently, in \cite{duran_stabilizing_2018, duran_bifurcation_2019}, a generalized RED-based model was proposed in which two new control parameters are introduced by means of a nonlinear packet dropping probability function, namely the normalized incomplete beta function. In addition, in \cite{m_amigo_generalized_2020,amigo_internet_2021} a theoretical analysis of the global stability was performed, and the results were used to find robust ranges of the new control parameters. This theoretical study has helped to design the new AQM algorithms proposed here for queue management, since they showed the existence of a wide range of parameters for which a higher stability than RED is achieved. One of the advantages of the algorithms proposed here is that the parameters are easily and intuitively adjusted with guarantees of achieving a good balance between stability and performance, and supported by the theoretical analysis made in \cite{m_amigo_generalized_2020,amigo_internet_2021}.

The idea of using a nonlinear packet drop probability function also appears in several papers. In \cite{feng_congestion_2017} the authors designed an algorithm called Three-section Random Early Detection (TRED), where the probability function is divided into three sections in order to achieve a trade-off between delay and throughput by distinguishing between light, moderate and high loads. Following a similar idea, an AQM scheme called Flexible Random Early Detection (FXRED) is proposed in \cite{vishnevskiy_flexible_2020} where different variations of the packet drop probability function are applied according to the state of the network traffic load. Also, in \cite{patel_new_2019}, a new probability model with a variation of the packet drop function with respect to RED is proposed, obtaining an increase in throughput and a reduction in the expected end-to-end delay.

There is a huge amount of research on the design of new dynamical AQM algorithms whose goal is to maintain a stable average queue length around a predetermined target under changing network traffic conditions. In fact, we can find recent work addressing the issue that show that the problem is still open. In \cite{xiong_novel_2008,bhatnagar_stochastic_2018, sharma_p-red_2018,abu-shareha_enhanced_2019} the RED algorithm is modified in order to improve its performance and stability for various network states, obtaining better results compared to their predecessors. In \cite{baklizi_weight_2020} the \textit{Weight Queue Active Queue Management} (WQDAQM) scheme, based on dynamic monitoring and reaction depending on the traffic load, is proposed. This algorithm aims to maintain the average queue length between two dynamically predetermined thresholds to prevent the buffer from exceeding the latter and overflowing.

For our numerical comparison in Sections \ref{subsec-beta-red-simulations} and  \ref{subsec-dynamic-beta-red-simulations} we have chosen some representative AQM algorithms that are implemented in ns-3, such as \textit{Adaptive RED} (ARED), \textit{Control Delay} (CoDel) and \textit{Proportional Integral Controller
Enhanced} (PIE). We add next a brief summary of the most important features of these algorithms.

ARED, proposed by Feng et al. \cite{feng_self-configuring_1999} and subsequently improved by Floyd et al. in \cite{floyd_adaptive_2001}, was intended to adjust the RED parameters adaptively to achieve a queue length around a prefixed target queue length. The main parameter that is tuned is the maximum packet drop probability, for which it uses an \textit{Additive-Increase-Multiplicative-Decrease} (AIMD) approach. However, the performance of ARED is inferior to that of RED when faced with complex network environments.

CoDel \cite{nichols_controlling_2012,nichols_controlled_2018} manages congestion control through the time that packets are in a given buffer (\textit{sojourn time}), or the time a given packet spends in the queue of a buffer. Thus, CoDel distinguishes between a ``good queue'', one that does not show bufferbloat, maintaining an adequate delay in the face of bursts of traffic, and a ``bad queue'', one that, on the contrary, has a high buffer delay in the face of low utilization.

PIE (see \cite{pan_pie_2013,pan_proportional_2017}) uses an estimate of the buffer queue delay as an indicator of congestion, marking with this estimated time each packet at the buffer entrance. When queuing a packet, a random discard is performed with a probability $p$ obtained as a function of the latency calculated as an estimate of the delay and the trend that this value develops. PIE adopts the model \textit{Proportional Integral} (PI) \cite{hollot_designing_2001} to maintain the queuing delay at a specified target value. Furthermore, the algorithm self-adjusts the control parameters as a function of the level of congestion, directly reflecting this measure in the current discard probability, this being updated at regular intervals.

The CoDel and PIE algorithms are designed with the main objective of minimizing queue latency while maintaining high link utilization. They represent solutions to the problem raised in \cite{f_baker_ietf_2015} of the \textit{Internet Engineering Task Force} (IETF), where a call for the design of new methods to control network latency is made. For dumbell topologies, both algorithms have performed well and mainly serve the purpose for which they were designed, which is to allow packet bursts to fill the buffer queue, preventing the queue from stalling under a persistent packet load.

\section{Scenarios of simulation and metrics} \label{sec-scenarios-metrics}

The complexity of the network means that the mathematical models of the protocols are not completely realistic and often the theoretically optimal algorithms do not perform as expected in real networks. This inevitably leads to the use of numerical simulations to obtain reliable predictions about the behavior of AQM algorithms in a real environment.  Furthermore, in most cases (e.g., for TCP), the congestion control algorithm for a given transport protocol is implemented in the same code base as the core of that protocol and is therefore the same for each end-to-end connection. Therefore, it is not possible to customize the response of the congestion algorithm to the characteristics of each connection.

\subsection{Scenarios of simulation}\label{subsec-scenarios-simulation}

\begin{figure}[htbp]
	\centering
	\includegraphics{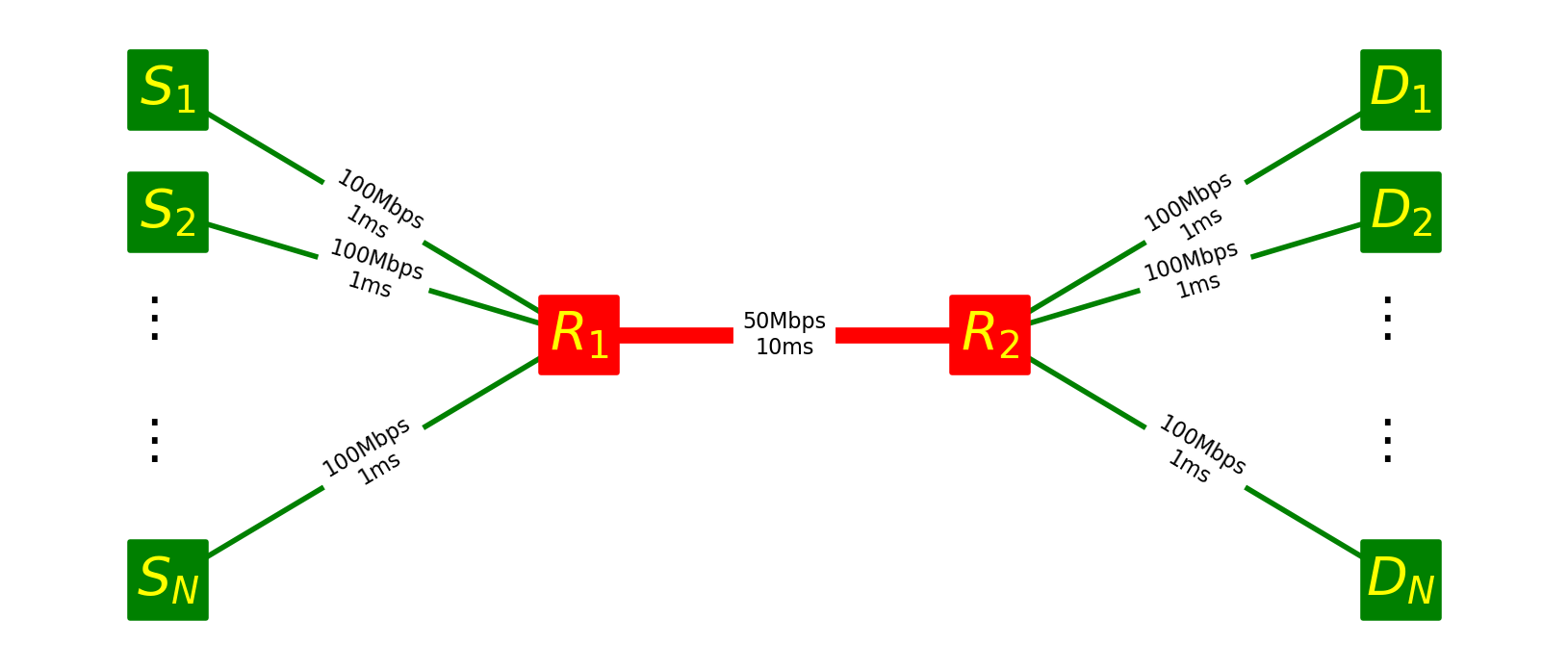}
	\caption{Network topology}
	\label{fig-dumbbell-topology}
\end{figure}

For testing the AQM algorithms, the topology used for the simulation (Sections \ref{subsec-beta-red-simulations} and  \ref{subsec-dynamic-beta-red-simulations}) is a simple dumbbell topology (see Figure \ref{fig-dumbbell-topology}), where $N$ is the number of long-lived TCP connections sharing a single bottleneck. Traffic is generated from the left side to the right side, specifically for each $i\in \left\{1,2,\ldots,N\right\}$, $S_i$ and $D_i$ denote the source and destination of the TCP flow $i$. The router $R_1$ on the left is where the bottleneck is actually located and the control AQM model will be installed. All other nodes, by default, have the Drop Tail queue manager installed. The edge links between the TCP sources and the router $R_1$, and the router $R_2$ and the TCP sinks have a capacity of $100\,\mathrm{Mbps}$ with a mean of $1\,\mathrm{ms}$ propagation delay. Router $R_1$ is connected to $R_2$ through a capacity $C$ of $50\,\mathrm{Mbps}$ and $10\,\mathrm{ms}$ propagation delay. The maximum buffer size $B$ of each router is set to $1000$ packets of a size $M$ of $1000$ bytes each. The TCP transport agent will be TCP Cubic.  

The \textit{level of congestion} of the simulation is set according to the number $N$ of long-lived TCP connections sharing the bottleneck. In that sense we distinguish two different scenarios:
\begin{description}
	\item[Scenario 1: ] Constant congestion level. In this case the number $N$ of active flows is constant throughout the simulation time, which is of $250$ seconds. 
	\item[Scenario 2: ] Changing congestion level. In this case the number $N$ of active flows varies as a function of time according to the following distribution: 
	\begin{itemize}
		\item Between $0$ and $50$ seconds the number of active flows is $N=100$.
		\item Between $50$ and $100$ seconds the number of active flows is $N=200$. 
		\item Between $100$ and $150$ seconds the number of active flows is $N=N_{\max}$. 
		\item Between $150$ and $200$ seconds the number of active flows is $N=200$. 
		\item Between $200$ and $250$ seconds the number of active flows is $N=100$. 
	\end{itemize}
	The congestion level in this case will be determined by the maximum number of active flows $N_{\max}$ in the simulation. 
\end{description}

\subsection{Performance metrics}\label{subsec-performance-metrics}

The ultimate goal of TCP/AQM protocols is to improve the performance of end-user metrics. However, the final performance is strongly influenced by the performance of router-based metrics. Thus, to evaluate the performance of the proposed algorithms, we have chosen to use two router-based metrics and three end-user metrics. The following is a brief description of the most important characteristics of the metrics used for performance evaluation.

\begin{itemize}

\item \textit{Average queue length} (AQL). The queue size indicates the number of packets pending transmission in the buffer queue. An unstable system is usually characterized by a synchronization in the TCP queues, accompanied by strong oscillations. 

Under suitable settings, the average queue length tends to stabilize at a value that we refer to as \textit{the equilibrium point}. One of the main objectives of AQM algorithms is to stabilize the buffer queue size and, in this sense, we shall refer to \textit{the stability of an AQM algorithm} as the ability to maintain the average queue length (or the average queuing delay) around a certain target value. The stability is an important performance characteristic of TCP/IP networks.

\item \textit{The packet drop rate.}
It measures the ratio of the number of packets dropped by an AQM to the total number of packets in queue. In this count, packets dropped by the link or channel at the physical layer are not counted, considering only the drop rate at the network layer. The main objective of an AQM is to maintain a stable queue size with as low packet drop rate as possible. This increases the performance, since dropped packets are an early signal of congestion to the TCP, causing a decrease in its send rate.

\item \textit{End-to-end throughput.} This is a performance measure obtained between two interlocutors (server-host). It measures the actual transmission of the total data propagated with respect to the simulation time (from the time the data is sent to the time it is received). It is defined as the number of bits received correctly per unit of time. Specifically, the calculation of this metric is obtained from the ratio between the number of bits received by the server/host and the time elapsed between the reception of the first segment and the last one. To calculate the throughput in the dumbbell type topology, we have averaged the ratio between the number of bits received by the hosts on the right, and the sum of elapsed time between the reception of the first segment and the last one at each of these nodes.

\item \textit{End-to-end delay (latency)}. It is one of the most significant metrics in a communication system and, in general terms, it is the time required to transmit a segment along its entire length, end-to-end. Specifically, it is calculated using the equation:
\begin{equation*}
	\text{Latency} = \text{propagation time} + \text{transmission time} + \text{queuing time} + \text{processing delay}
\end{equation*}
For its calculation in the dumbbell topology, the end-to-end delay times of all the segments sent between the left and right hosts are summed, divided by the number of segments received on both hosts.

\item \textit{Jitter}. Jitter in a flow is defined as the variation in delay of arriving packets over time. A very high jitter can cause packet loss due to buffer overflow. In the dumbbell topology, an average value is calculated by summing the time variations between the correlative packets of all flows, and dividing by the number of variations.

\end{itemize}

All simulations presented in this paper were performed with ns-3 \cite{nsnam_network_2011}. This is a discrete event network simulator for Internet systems, primarily intended for research and educational use. It is free software, licensed under GNU GPLv2, and is publicly available for research, development and use.

\section{The Beta RED scheme} \label{sec-beta-red}

The Beta RED algorithm that we introduce is inspired by the classical RED, where the packet drop probability function is dependent on two new parameters. The idea is to adapt the parameters  in order to improve the performance and stability of the system according to the characteristics of the congestion scenarios, such as traffic load, available bandwidth and desired delay, etc. At first it might be thought that introducing two new parameters complicates the already intricate parameter tuning that the RED algorithm undergoes since its inception, but we will show that it is actually an advantage, since it is possible to easily and intuitively adapt these parameters to stabilize the average queue length as close as possible to a preset reference value. We will now briefly describe the biparametric family of probability beta functions and the properties on which we will rely in the design of the new algorithm.

\subsection{Normalized incomplete beta function}

The beta distribution function (or normalized incomplete beta function) $I(\alpha,\beta)$ is given by the expression
\begin{equation}
I_z(\alpha,\beta)=\frac{\mathfrak{B}(z;\alpha ,\beta )}{\mathfrak{B}(1;\alpha ,\beta )},\;\;\;\;\mathfrak{B}(z;\alpha ,\beta
)=\int_{0}^{z}t^{\alpha -1}(1-t)^{\beta -1}dt,  \label{I(z)}
\end{equation}
with $\alpha ,\beta >0$ and $z\in[0,1]$. By definition, $I_z(\alpha,\beta)$ is strictly increasing, hence invertible. Its inverse, $I^{-1}_z(\alpha ,\beta)$, is also strictly
increasing.

Although usually the family of probability beta functions are described according to the parameters $\alpha$ and $\beta$, here it will be much more convenient to describe them according to their mean $\mu$ and standard deviation $\sigma$, as we will show later. It is well known that the expected value and the variance of the beta distribution is given by:
\begin{equation}\label{eq-mu-sigma}
	\mu=E[I(\alpha,\beta)]=\frac{\alpha}{\alpha+\beta},\qquad \sigma^2=Var[I(\alpha,\beta)]=\frac{\alpha\beta}{(\alpha+\beta)^2(\alpha+\beta+1)}.
\end{equation}
If we solve $\alpha$ and $\beta$ as a function of $\mu$ and $\sigma^2$ in the above equations we obtain:
\begin{equation}
	 \alpha(\mu,\sigma)=\mu \left(\frac{\mu(1-\mu)}{\sigma^2}-1\right),\qquad  \beta(\mu,\sigma)=(1-\mu) \left(\frac{\mu(1-\mu)}{\sigma^2}-1\right).
\end{equation}
Thus, given values $\mu\in(0,1)$ and  $\sigma<\sqrt{\mu(1-\mu)}$, we have univocally determined parameters  $\alpha$ and $\beta$ such that the beta distribution $I(\alpha,\beta)$ has mean $\mu$ and standard deviation $\sigma$. Henceforth we will consider the beta distribution $\tilde{I}(\mu,\sigma)$ with respect to the parameters $\mu$ and $\sigma$, specifically:
\begin{equation}
	\tilde{I}_z(\mu,\sigma)=I_z(\alpha(\mu,\sigma),\beta(\mu,\sigma))
\end{equation}

\subsection{Beta RED algorithm}

In \cite{duran_stabilizing_2018,duran_bifurcation_2019} the authors generalize the RED scheme by replacing the linear packet drop probability function  $p_{\max}\cdot z(q_{\mathrm{avg}})$ in \eqref{p-ave} by the nonlinear function $p_{\max}\cdot I_{z(q_{\mathrm{avg}})}(\alpha,\beta)$, where $0\leq z\leq 1$ and $\alpha,\beta >0$. Since $I_z(1,1)=z$, we recover the classical RED scheme for $\alpha =\beta =1$. The purpose of this generalization was to improve the stability properties by introducing the additional control parameters $\alpha $ and $\beta$. Following the analysis of Ranjan \cite{ranjan_nonlinear_2004}, in \cite{m_amigo_generalized_2020,amigo_internet_2021} the authors perform a detailed theoretical study of chaos, bifurcation diagrams, Lyapunov exponents and global stability robustness for different control parameters and fixed system parameters. Bifurcation diagrams are discussed for specific values of $\alpha$ and $\beta$ in different scenarios, as well as biparametric sweeps of these parameters in which it was found that there are parameter regions in which the system performs very successfully in terms of stability and robustness.

However, for the design of our algorithms it is much more convenient to consider the biparametric family of packet drop probability functions with respect to the parameters $\mu$ and $\sigma$, since they have an intuitive meaning in statistical terms, namely, the mean and the standard deviation. In this case the packet drop probability function takes the expression
\begin{equation}  \label{beta-p-ave}
	p(q_{\mathrm{avg}})= 
	\begin{cases}
	0 & \text{if }q_{\mathrm{avg}}<q_{\min }, \\ 
	1 & \text{if }q_{\mathrm{avg}}>q_{\max }, \\ 
	p_{\max } \cdot  \tilde{I}_{z(q_{\mathrm{avg}})}(\mu,\sigma) & \text{otherwise.}\end{cases}
\end{equation}
where $\mu\in(0,1)$ and $\sigma<\sqrt{\mu(1-\mu)}$. From Equation \eqref{eq-mu-sigma}, the classical RED ($\alpha=\beta=1$) is recovered when $\mu=\frac{1}{2}=0.5$ and $\sigma=\frac{1}{2\sqrt{3}}\approx 0.2886$.

\begin{figure}[htbp]
	\centering
	\includegraphics{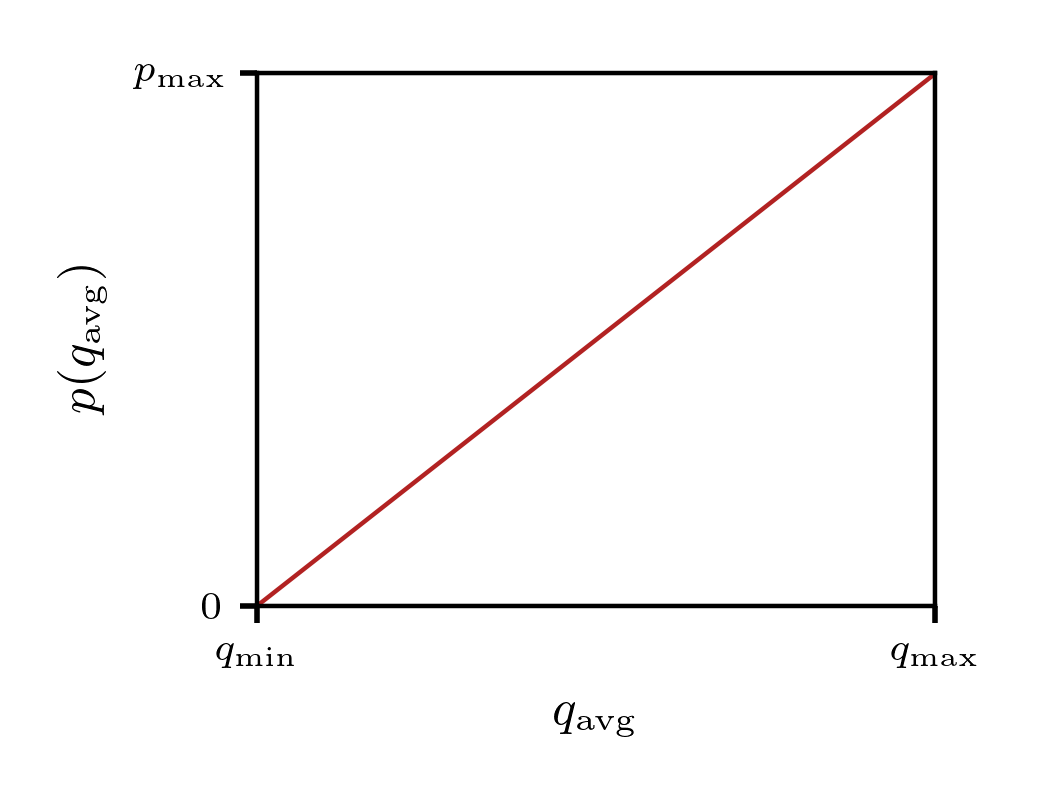}
	\includegraphics{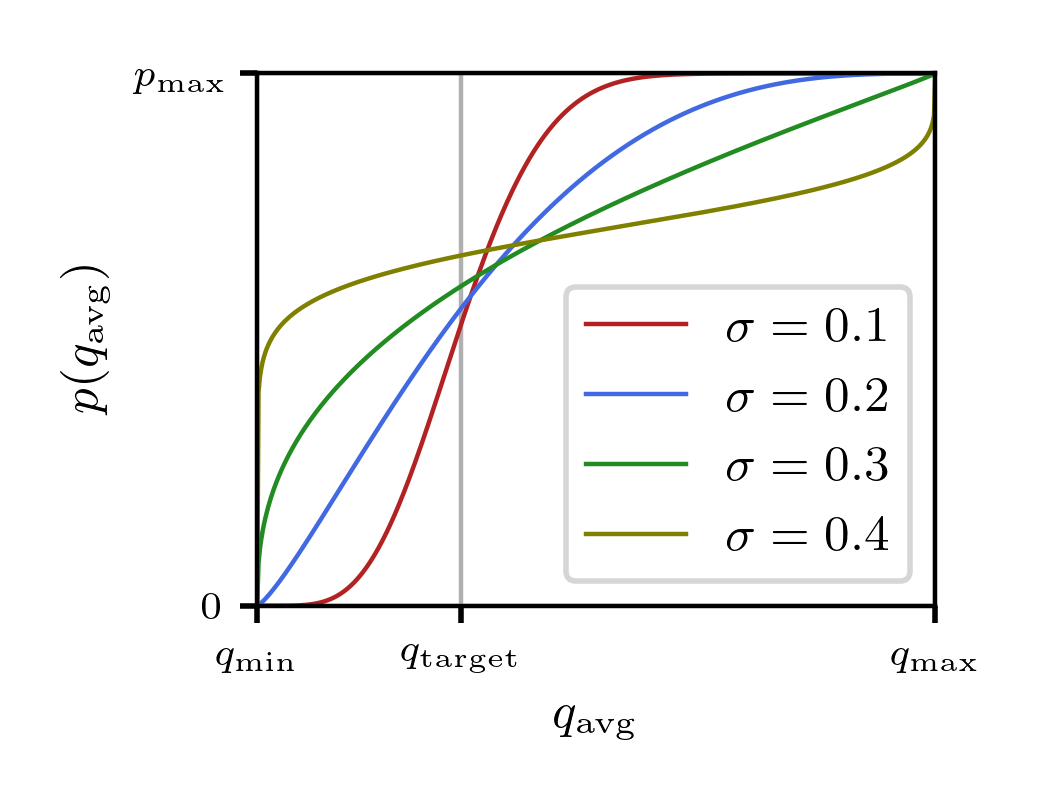}

	\caption{The lower the standard deviation $\sigma$, the higher the concentration of the packet drop probability mass around $q_{\mathrm{target}}$.}
	
	\label{fig-drop-probability}
\end{figure}

As in the RED algorithm, the basic idea of BetaRED is to maintain the average queue length $q_{\mathrm{avg}}$ (calculated through the EWMA algorithm) within minimum and maximum thresholds $q_{\min}$ and $q_{\max}$, but as close as possible to a predetermined target queue length that we denote by $q_{\mathrm{target}}$. With this objective in mind, the next step is to provide concrete values of $\mu$ and $\sigma$ for which a good balance between stability and performance of the algorithm is obtained. From this point of view, our selection of the mean is given as a function of the target queue length $q_{\mathrm{target}}$ and will be completely determined by means of the expression:
\begin{equation}
	\mu=\frac{q_{\mathrm{target}}-q_{\min}}{q_{\max}-q_{\min}}.
\end{equation}
With this choice we match the mean of the packet drop probability function to the $q_{\mathrm{target}}$ value. Moreover, the standard deviation $\sigma$ is a spread measure of the values of the distribution around $\mu$. This means that the smaller $\sigma$ is, the more concentrated the packet drop probability mass is around its mean $\mu$ (see Figure \ref{fig-drop-probability}). Therefore, when the buffer starts to fill up and the average length of the buffer queue $q_{\mathrm{avg}}$ approaches $q_{\mathrm{target}}$, the probability of packet drop will increase faster the smaller $\sigma$ is. Moreover, the smaller $\sigma$ is, the closer to $q_{\mathrm{target}}$ the system will stabilize. Consequently, the value of the standard deviation is a parameter that has to be adjusted in our BetaRED scheme. However, the fact that the value of the standard deviation verifies $0<\sigma<\sqrt{\mu(1-\mu)}=\sigma_{\max}$ causes a slight inconvenience since the user has to calculate the maximum value $\sigma_{\max}$ before selecting the value of $\sigma$. To avoid this inconvenience, the $\sigma$ value is selected by mean of a scale factor $\theta$ that verifies $0<\theta<1$, being $\sigma=\theta\cdot\sqrt{\mu(1-\mu)}$, which facilitates the task to the user. Summarizing, in BetaRED, once the target queue length $q_{\mathrm{target}}$ is set, we have to adjust the parameters $q_{\min}$, $q_{\max}$, $p_{\max}$, $w$ and $\theta$. The rest of the algorithm works exactly the same as RED. For the sake of completeness, we present an outline of the pseudo-code for BetaRED in Algorithm \ref{alg-beta-red}.

\begin{algorithm}[htbp]
\caption{Pseudo-code outline for the Beta RED algorithm.}
\label{alg-beta-red}
\setstretch{1.2}
\begin{algorithmic}[1]
   \State set tunable parameters: $q_{\mathrm{target}}, q_{\min}, q_{\max}, w, p_{\max}, \theta$
    \State $\displaystyle \mu=\frac{q_{\mathrm{target}}-q_{\min}}{q_{\max}-q_{\min}}; \quad \sigma=\theta\cdot\sqrt{\mu(1-\mu)};$
   \For{each arriving packet}
	   \State calculate new $q_{\mathrm{avg}}=(1-w)\cdot q_{\mathrm{avg}}+w\cdot q_{\mathrm{cur}}$
	   \If{$q_{\mathrm{avg}} \leq q_{\min}$}
	   	\State $p=0$ 
	   \ElsIf{$q_{\mathrm{avg}} \geq q_{\max}$}
	    \State $p=1$
	   \Else
	    \State $p=p_{\max } \cdot  \tilde{I}_{z(q_{\mathrm{avg}})}(\mu,\sigma)$
	   \EndIf
	   \State with probability $p$, \textbf{drop arriving packet}
   \EndFor
\end{algorithmic}
\end{algorithm}

\subsection{Simulations for the BetaRED algorithm}
\label{subsec-beta-red-simulations}

In this section we present some numerical simulations with the purpose of analyzing the average queue length and the performance of the BetaRED algorithm for some values of the control parameters. However, space limitation conditions us to present only a selection of a larger set of the simulations that have been carried out. The analysis of the results obtained will help us to design two new dynamic algorithms based on BetaRED, which will be described in Section \ref{sec-dynamic-beta-red}.

\begin{figure}[htbp]
	\centering
	\includegraphics{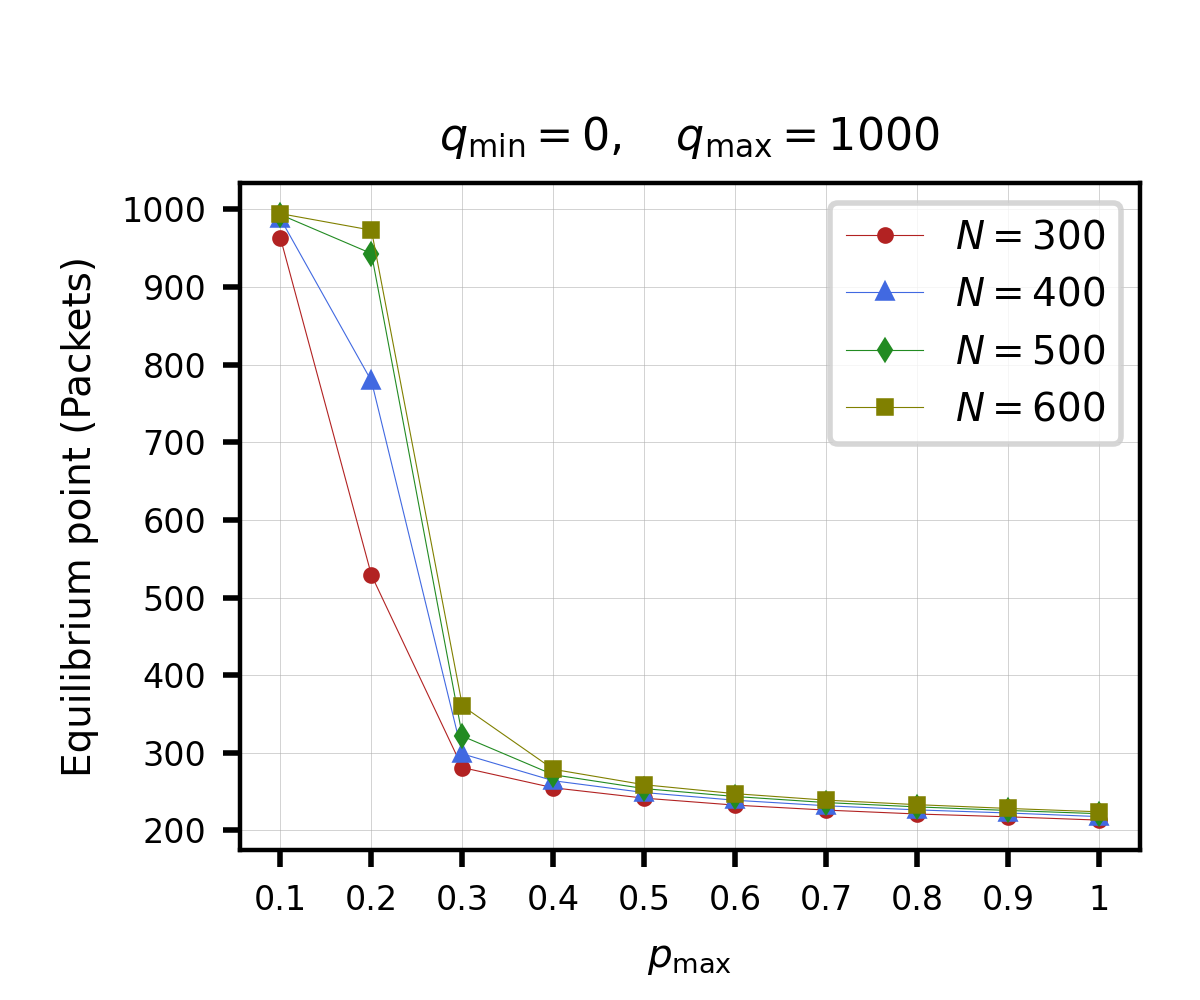}
	\includegraphics{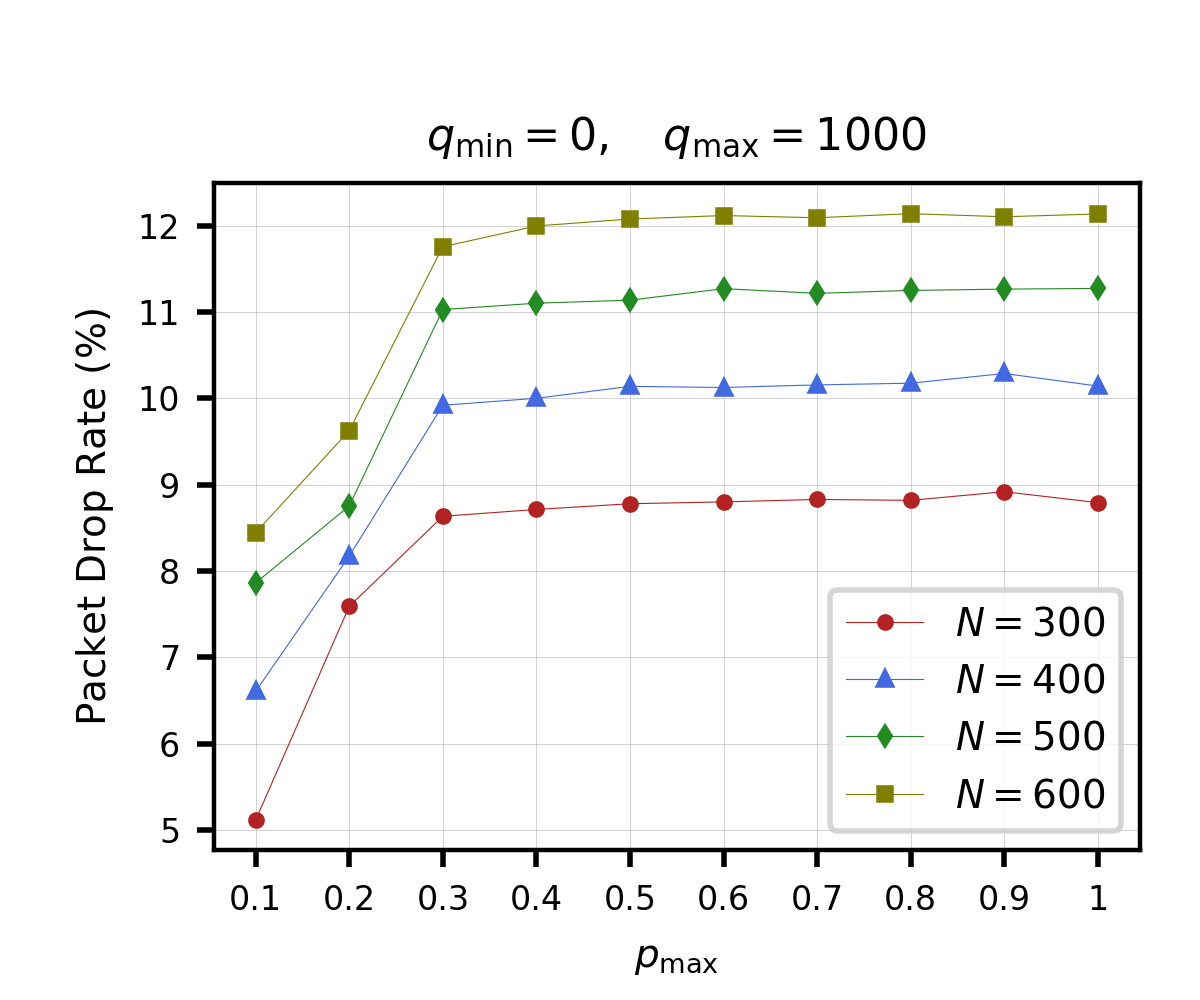}
	\includegraphics{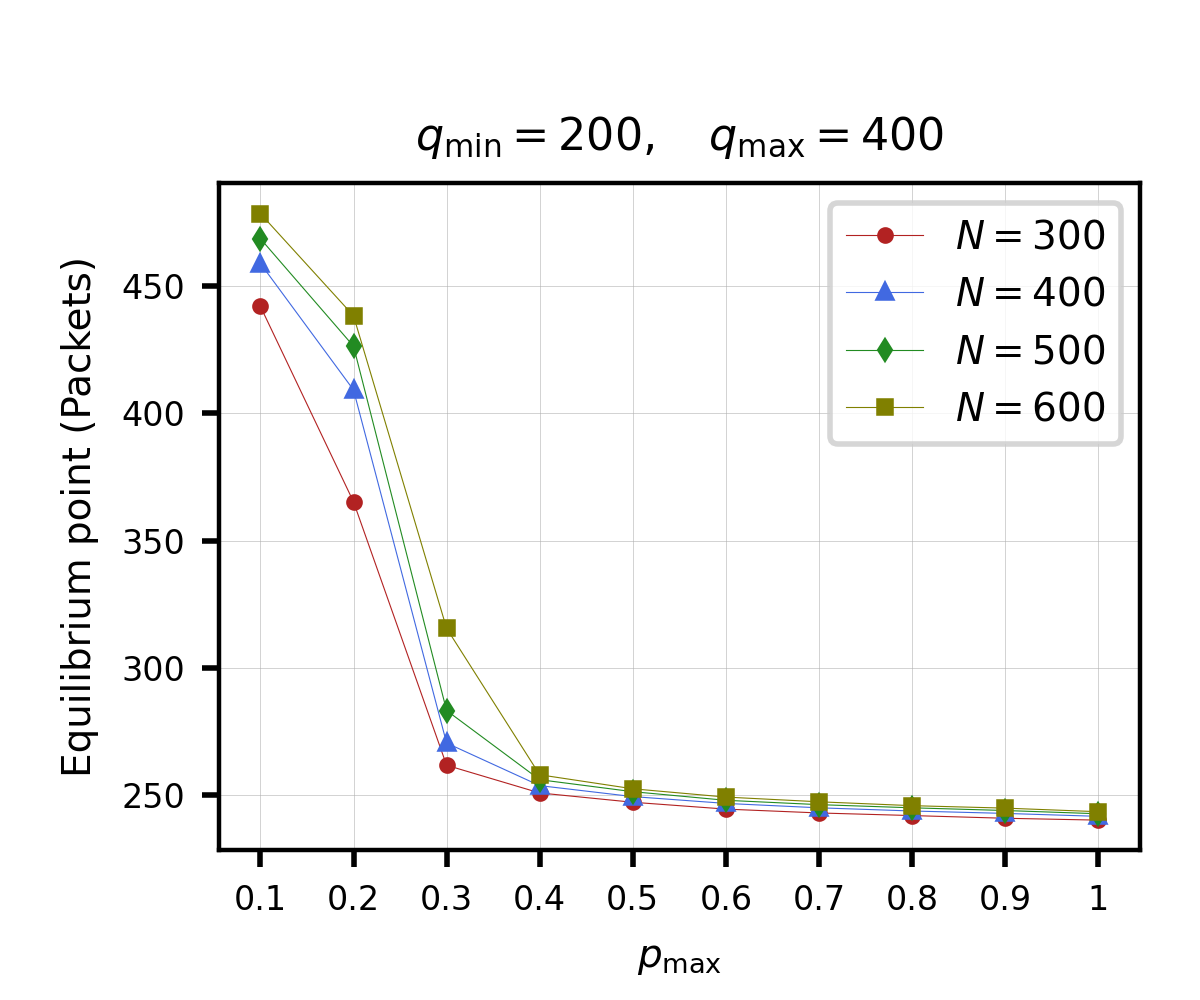}
	\includegraphics{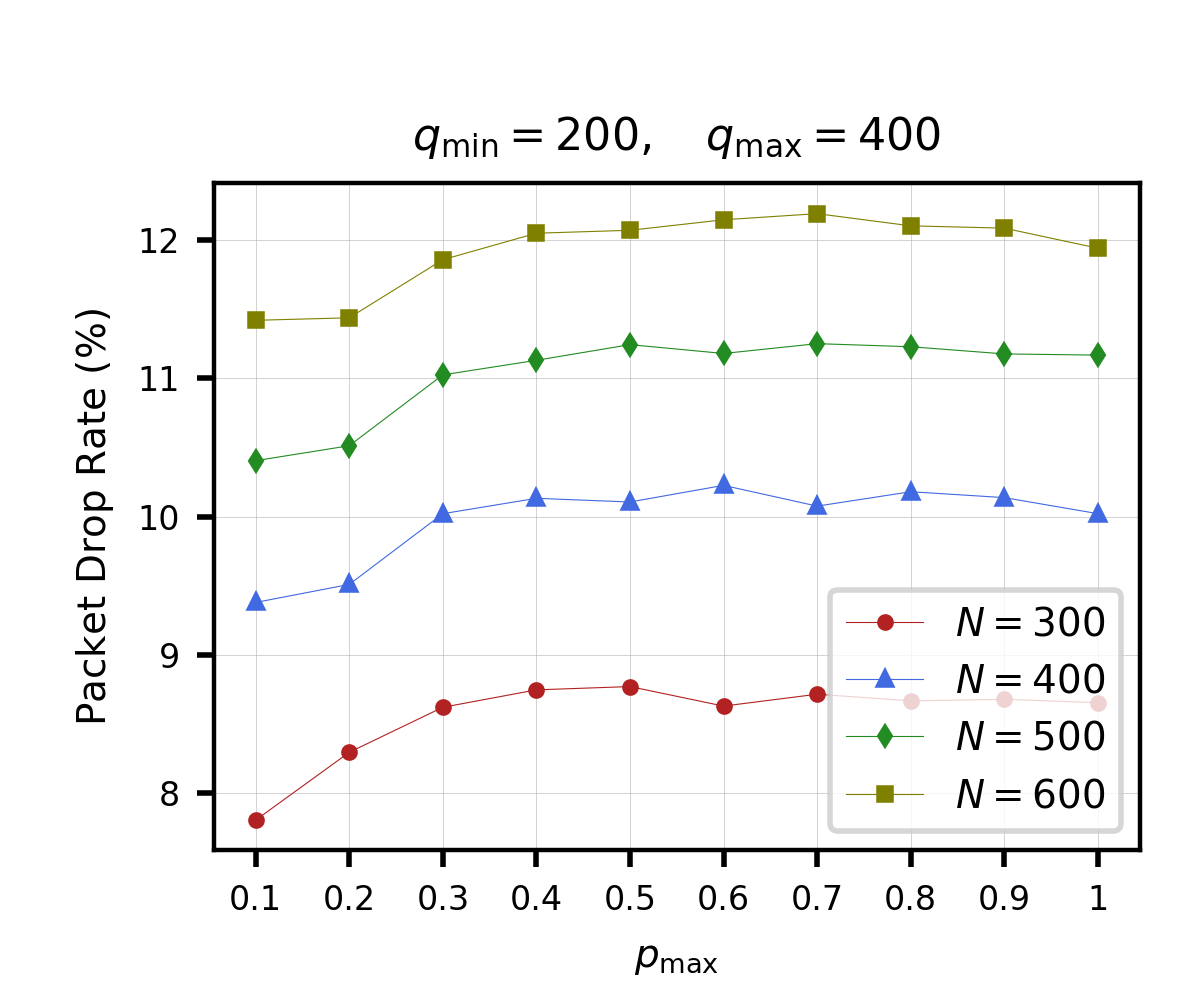}

		\caption{Comparison of the BetaRED algorithm according to the number $N$ of FTP active flows and different values of the maximum probability threshold $p_{\max}$. Different thresholds $q_{\min}$ and $q_{\max}$ are also considered. Other tunable parameters:  $q_{\mathrm{target}}=250\,\mathrm{packets}$, $w=0.1$, $\theta=0.1$. Scenario 1 on the dumbbell topology (described in Section \ref{subsec-scenarios-simulation}) is used. As $p_{\max}$ and/or the level of congestion increase, the equilibrium point of the average queue length gets closer to the prefixed $q_{\mathrm{target}}$. On the other hand, if minimum and maximum thresholds $q_{\min}$ and $q_{\max}$ are closer to the target queue length $q_{\mathrm{target}}$, this also results in the average queue length equilibrium point being closer to $q_{\mathrm{target}}$.}
	
	\label{fig-beta-red-pmax-N}
\end{figure}

In the first set of simulations (see Figure \ref{fig-beta-red-pmax-N}) we vary the parameter $p_{\max}$ for different numbers of nodes in the dumbbell topology shown in Figure \ref{fig-dumbbell-topology}, and analyze the average queue length in the last $125$ seconds (half of the total simulation time). We contemplate only the second half of the total simulation  considering the first half as a transition period, and thus obtaining a better estimate of the equilibrium point. Two different scenarios are setting according to the parameters $q_{\min}$ y $q_{\max}$. The first one when $q_{\min}=0$ and $q_{\max}=1000$ (maximum buffer capacity). In the second scenario we consider the minimum and maximum thresholds close to the target queue length $q_{\mathrm{target}}=250$, namely, $q_{\min}=200$ and $q_{\max}=400$.

We note that to achieve a higher robustness in terms of the stability of the average queue length it is convenient to select the value $p_{\max}=1$. This way, the average queue length stays closer to $q_{\mathrm{target}}$. However, the performance may decrease for increasing values of the parameter $p_{\max}$, which means that we should look for the lowest possible $p_{\max}$ parameter value that guarantees stability and increases performance. It is well known that the ARED algorithm \cite{floyd_adaptive_2001} is based precisely on adjusting $p_{\max}$ to obtain a predetermined target average queue length. In Section \ref{sec-dynamic-beta-red} we will follow the same idea of ARED adapted to BetaRED.

\begin{figure}[htbp]
	\centering
	\includegraphics{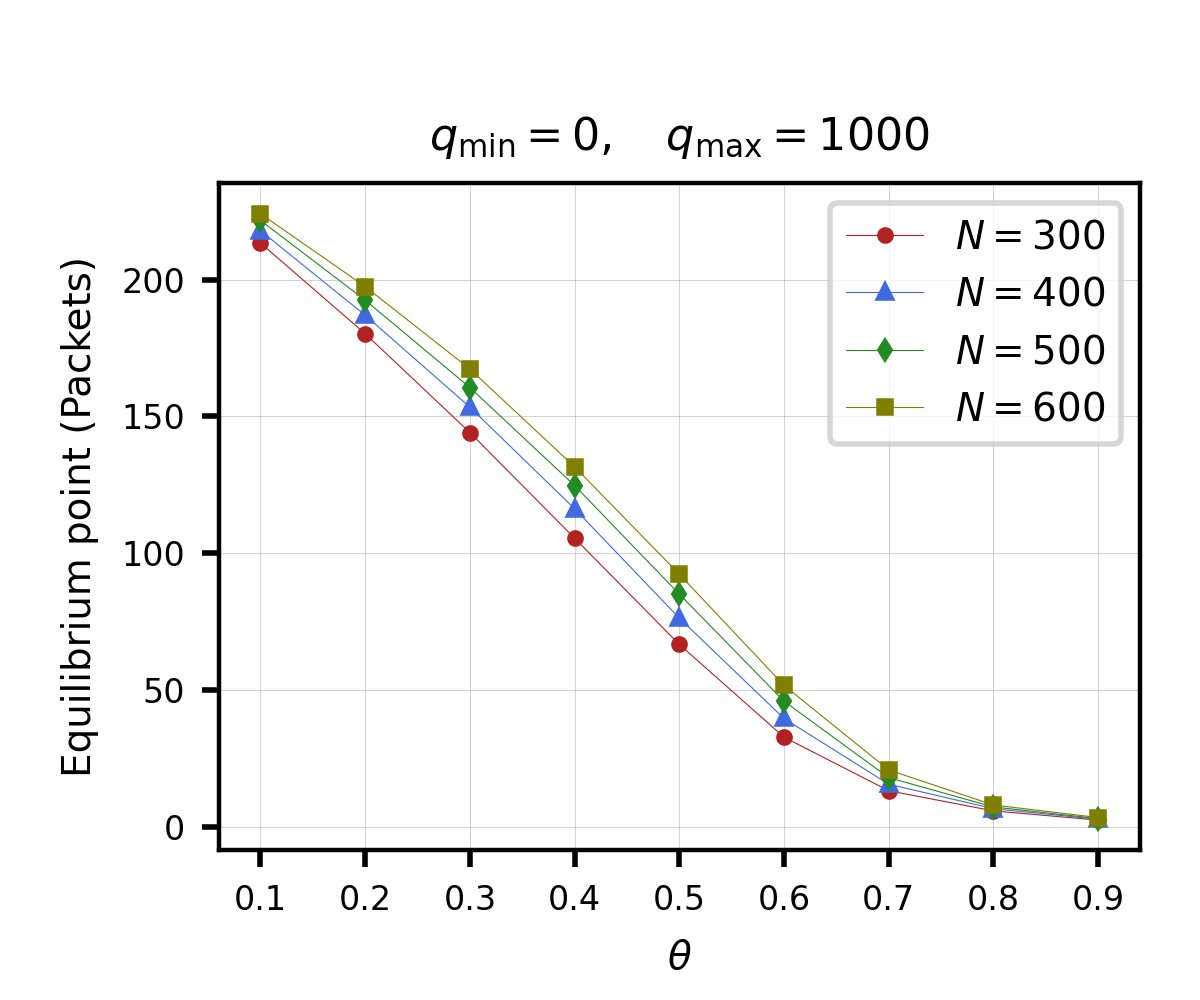}
	\includegraphics{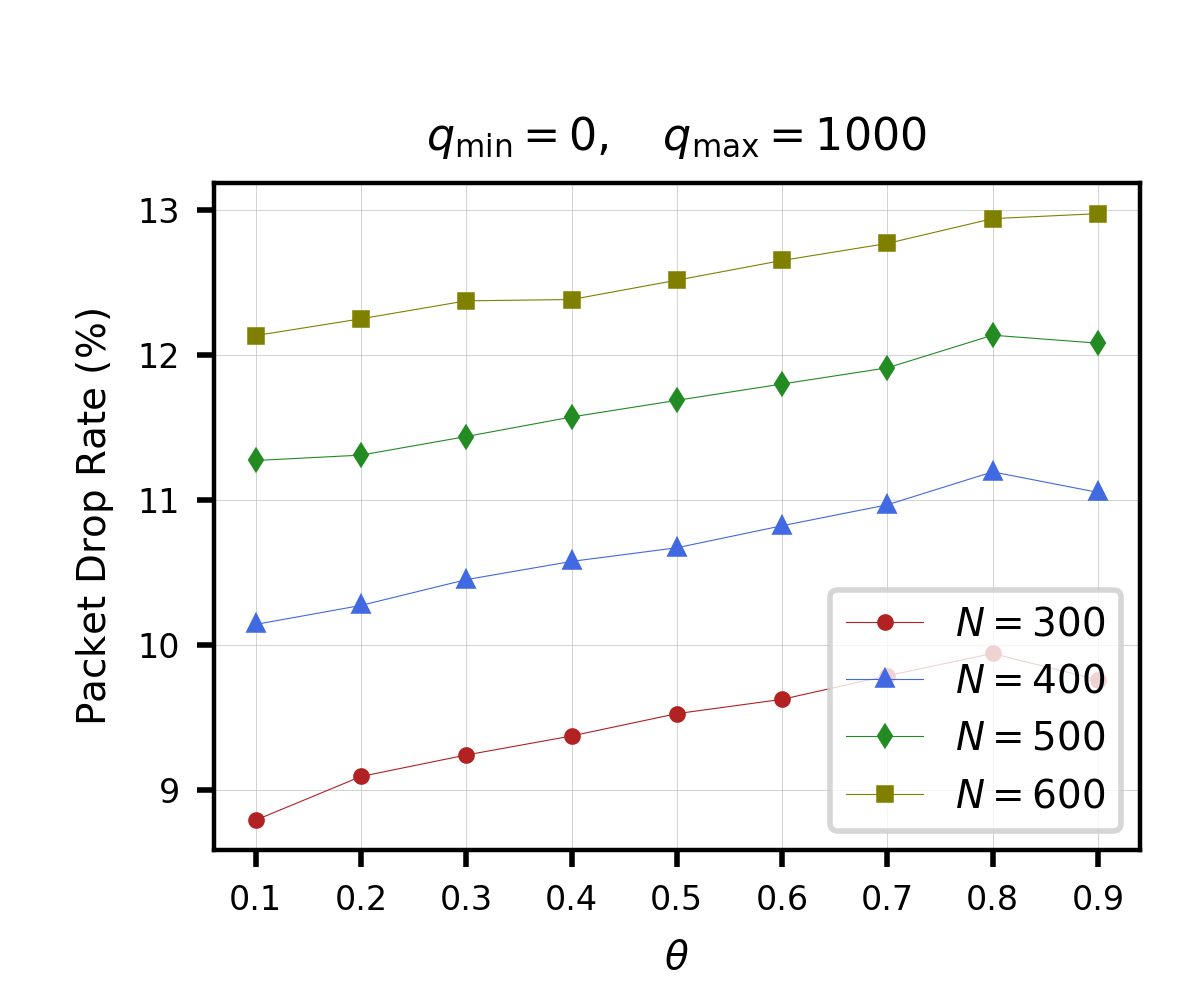}
	\includegraphics{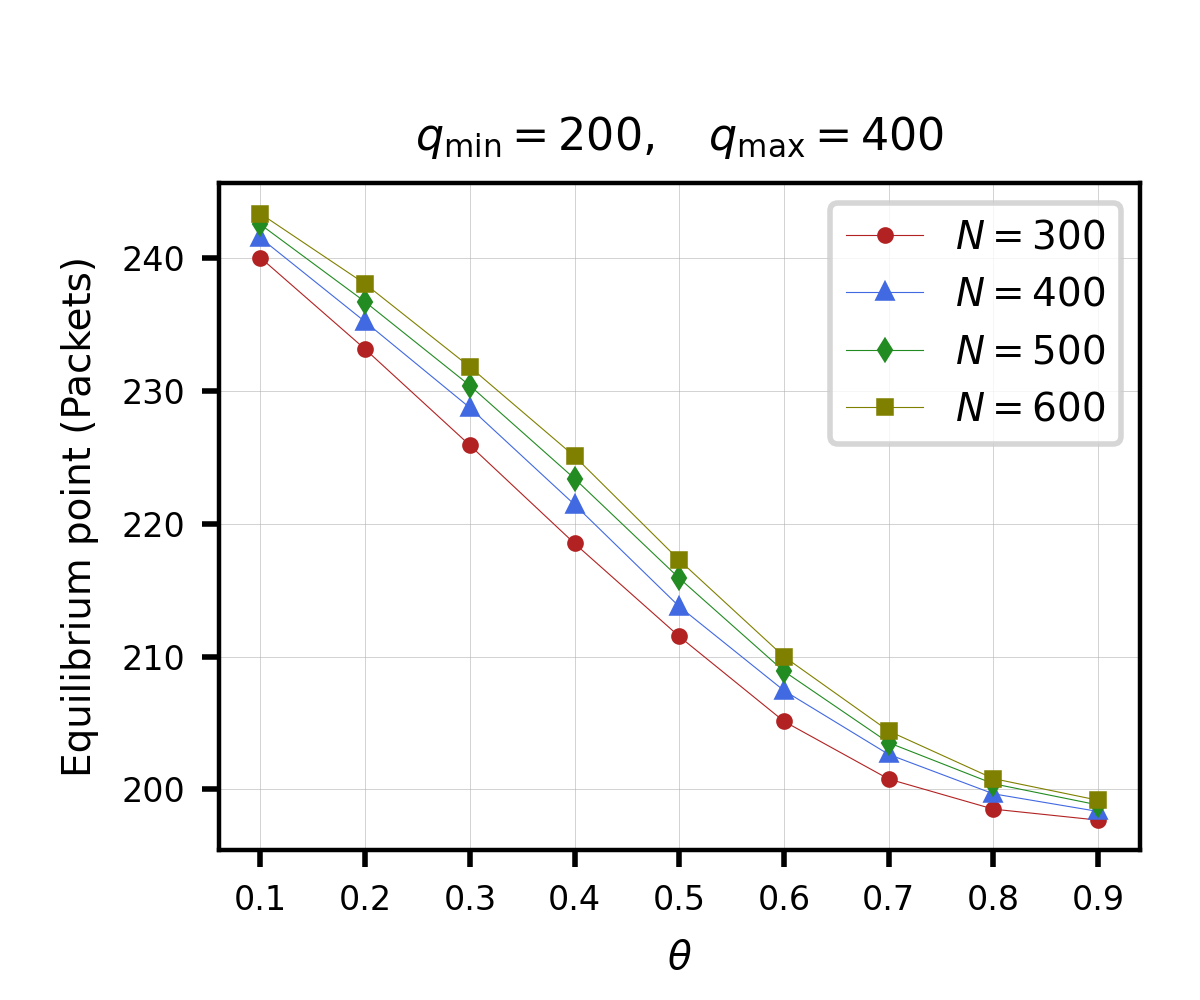}
	\includegraphics{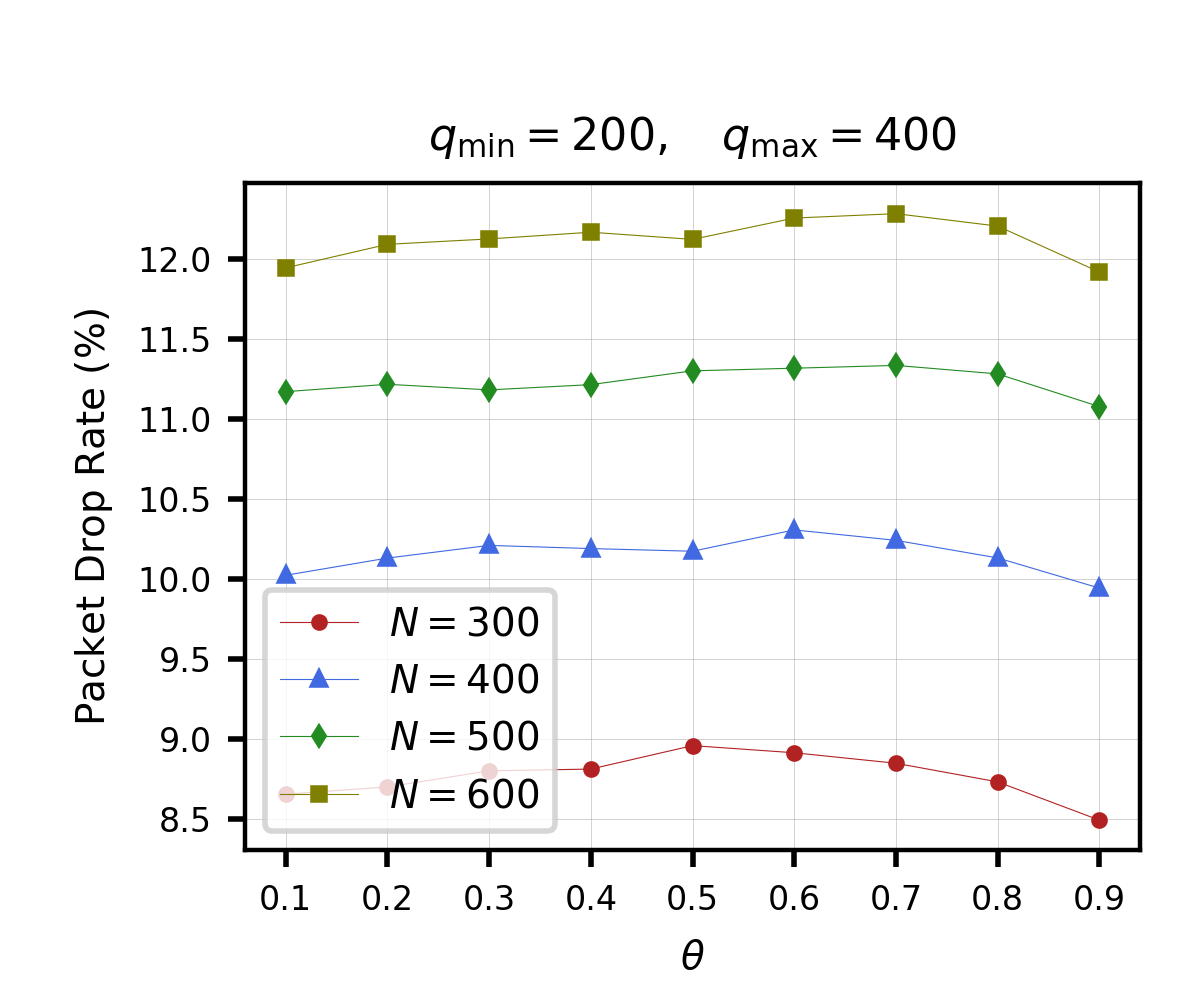}

		\caption{Comparison of the BetaRED algorithm according to the number $N$ of FTP active flows and different values of $\theta$. Different thresholds $q_{\min}$ and $q_{\max}$ are also considered. Other tunable parameters:  $q_{\mathrm{target}}=250\,\mathrm{packets}$, $w=0.1$, $p_{\max}=1$. Scenario 1 on the dumbbell topology (described in Section \ref{subsec-scenarios-simulation}) is used. As the standard deviation $\sigma$ decreases and the level of congestion increases, the equilibrium point of the average queue length gets closer to the prefixed $q_{\mathrm{target}}$. Similarly, if the minimum and maximum thresholds $q_{\min}$ and $q_{\max}$ are closer to the target queue length $q_{\mathrm{target}}$, then the average queue length equilibrium point is closer to $q_{\mathrm{target}}$.}
	
	\label{fig-beta-red-sigma-N}
\end{figure}

In a second set of simulations (see Figure \ref{fig-beta-red-sigma-N}), we vary the $\theta$ parameter for different numbers of nodes in the dumbbell topology presented in Figure \ref{fig-dumbbell-topology}. Again, to estimate the equilibrium point, we calculate the average queue length in the last $125$ seconds, i.e., in the second half of the total simulation time. As in the first set of simulations, we also consider the same two scenarios according to the minimum and maximum thresholds $q_{\min}$ and $q_{\max}$. As expected (taking into account the interpretation of the standard deviation $\sigma$), better stability results are obtained as the value of $\theta$ (and hence $\sigma$) decreases. In the following, the default value of $\theta=0.1$ will be used for simulations. 

The discussion of the results when we vary the value of the weight parameter $w$ is more intricate. Although we do not show concrete numerical simulations since the behavior is less predictable, it was observed that, as a general rule, increasing its value is accompanied by better performance, but stability gets worse. Nevertheless, we shall show that for the DBetaRED algorithm (described in Section \ref{subsec-dynamic-beta-red-simulations}) both high performance and good stability are obtained for a wide range of $w$ values.

\section{Dynamic algorithms based on BetaRED} \label{sec-dynamic-beta-red}

In the previous section we have designed a generic AQM based on the nonlinear beta probability distribution function. The major difficulty was, as in the case of RED and so many other schemes, the selection of the parameters to obtain good performance and stability around the target queue length. We found that the BetaRED algorithm tries to keep the average queue length close to the target queue length, but the level of congestion, the particular network conditions and the choice of appropriate values for the control parameters $q_{\min}$, $q_{\max}$, $p_{\max}$, $w$ and the  $\theta$, can cause that the average queue length at equilibrium point undergoes substantial deviations from the desired $q_{\mathrm{target}}$, as shown in Figures \ref{fig-beta-red-pmax-N} and \ref{fig-beta-red-sigma-N}.

In this section we are going to propose two dynamical schemes based on BetaRED that are more accurate in reaching the target queue length and adapt satisfactorily to network traffic changes. Specifically, the objective is to perform dynamic parameter adjustments that correct the deviation of the equilibrium value of $q_{\mathrm{avg}}$ from the $q_{\mathrm{target}}$ value. The possibilities of acting on the parameters to correct these deviations are diverse. We propose two: the first one follows the same approach as the ARED algorithm \cite{floyd_adaptive_2001}, which  is based on varying the value of $p_{\max}$ to correct the difference from the equilibrium point of $q_{\mathrm{avg}}$ to $q_{\mathrm{target}}$. The second is to modify dinamically the target queue length $q_{\mathrm{target}}$ to correct the deviations.

\subsection{The Adaptative Beta RED algorithm (ABetaRED)} \label{subsec-adaptative-beta-red}

ABetaRED is inspired by the ARED algorithm (which in turn is based on RED), although there are some relevant changes due to the differences between the RED and BetaRED algorithms. For our ABetaRED algorithm, the tuning of the following two parameters is straightforward:
\begin{equation}\label{eq-abeta-red-fixed-parameters}
	q_{\min}=0,  \quad  q_{\max}=B.  
\end{equation}
The choices for $q_{\min}$ and $q_{\max}$ are based primarily on the numerical simulations for BetaRED in Section \ref{subsec-beta-red-simulations}. 
It is observed that with the choices \eqref{eq-abeta-red-fixed-parameters} an acceptable performance is achieved for a very wide range of parameters in various scenarios, while no substantial improvement in performance was obtained with the selection of other values. The control parameters $\mathrm{Alpha}$ and $\mathrm{Beta}$  determine the size of the increase and decrease steps of the maximum probability, respectively, with the range of admissible values being $0<\mathrm{Alpha},\mathrm{Beta}\leq 1$.

On the other hand, the tunable parameters to be set by the user are the target delay $T_{\mathrm{target}}$, the scale factor $\theta$ and the averaging weight $w$. As in ARED, once the target delay is set, the target queue length can be estimated by
\begin{equation}
	\label{eq-estimated-qtarget}
	q_{\mathrm{target}}=C\cdot T_{\mathrm{target}}
\end{equation}
where, as in Equation \eqref{eq-abeta-red-fixed-parameters}, $C$ is the link capacity measured in packets per second, and $T_{\mathrm{target}}$ is measured in seconds. A pseudo-code for ABetaRED is outlined in Algorithm \ref{alg-adaptative-beta-red}.

\begin{algorithm}[htbp]
\caption{Pseudo-code outline for the Adaptative Beta RED algorithm. By default, $\mathrm{Alpha}=\mathrm{Beta}=1$ and $T_{\mathrm{update}}=0.5\,\mathrm{seconds}$, were set in all simulations.}
\label{alg-adaptative-beta-red}
\setstretch{1.2}
\begin{algorithmic}[1]
	\State set control parameters: $T_{\mathrm{target}}, w, \theta$
	\State $q_{\min}=0;\quad q_{\max}=B;\quad p_{\max}=0.5;\quad q_{\mathrm{target}}=C\cdot T_{\mathrm{target}}$;
	\State $\displaystyle \mu=\frac{q_{\mathrm{target}}-q_{\min}}{q_{\max}-q_{\min}}; \quad \sigma=\theta\cdot\sqrt{\mu(1-\mu)};$
	\For{each arriving packet}
		\State calculate new $q_{\mathrm{avg}}=(1-w)\cdot q_{\mathrm{avg}}+w\cdot q_{\mathrm{cur}}$
		\For{every interval time $T_{\mathrm{update}}$}
			\If{$q_{\mathrm{avg}} < q_{\mathrm{target}}$}
				\State decrease maximum probability: 
				\State $\displaystyle p_{\max}=\max \left[0.01\,,\,p_{\max}\cdot \mathrm{Alpha}\cdot \left(1-\frac{q_{\mathrm{target}}-q_{\mathrm{avg}}}{q_{\max}-q_{\min}}\right)\right]$
			\ElsIf{$q_{\mathrm{avg}} > q_{\mathrm{target}}$}
				\State increase maximum probability: 
				\State $\displaystyle p_{\max}=\min \left[0.99\,,\,p_{\max}+\mathrm{Beta}\cdot p_{\max}\cdot(1-p_{\max})\cdot\frac{q_{\mathrm{avg}}-q_{\mathrm{target}}}{q_{\max}-q_{\min}}\right]$
			\EndIf
		\EndFor 
		\State update $p=p_{\max} \cdot  \tilde{I}_{z(q_{\mathrm{avg}})}(\mu,\sigma)$
		\State with probability $p$, \textbf{drop arriving packet}
	\EndFor
\end{algorithmic}
\end{algorithm}

\subsection{The Dynamic Beta RED algorithm (DBetaRED)} \label{subsec-dynamic-beta-red}

The idea of the DBetaRED algorithm is, once the target queue delay  $T_{\mathrm{target}}$ is fixed and the target queue length is estimated as in ABetaRED by Equation \eqref{eq-estimated-qtarget}, we introduce another dynamic parameter called \textit{virtual target queue length} $\tilde{q}_{\mathrm{target}}$ to correct for deviations between the average queue length $q_{\mathrm{avg}}$ and the actual target queue length $q_{\mathrm{target}}$. According to the value of $\tilde{q}_{\mathrm{target}}$, the value of the mean $\mu$ (and thus the standard deviation $\sigma=\theta\cdot\sqrt{\mu(1-\mu)}$) of the drop probability function are updated dynamically for each time interval $T_{\mathrm{update}}$ previously established. 
An outline of the pseudo-code of this new algorithm is given in Algorithm \ref{alg-dynamic-beta-red}.

\begin{algorithm}[htbp]
\caption{Pseudo-code outline for the Dynamic Beta RED algorithm. By default, $\mathrm{Alpha}=\mathrm{Beta}=1$ and $T_{\mathrm{update}}=0.5\,\mathrm{seconds}$, were set in all simulations.}
\label{alg-dynamic-beta-red}
\setstretch{1.2}
\begin{algorithmic}[1]
	\State set control parameters: $T_{\mathrm{target}}, w, \theta$
	\State $q_{\min}=0; \quad q_{\max}=B; \quad  p_{\max}=1;$
	\State $q_{\mathrm{target}}=C\cdot T_{\mathrm{target}}$
	\State $\displaystyle \tilde{q}_{\mathrm{target}}=q_{\mathrm{target}}; \quad \mu=\frac{\tilde{q}_{\mathrm{target}}-q_{\min}}{q_{\max}-q_{\min}}; \quad \sigma=\theta\cdot\sqrt{\mu(1-\mu)};$
	\For{each arriving packet}
		\State calculate new $q_{\mathrm{avg}}=(1-w)\cdot q_{\mathrm{avg}}+w\cdot q_{\mathrm{cur}}$
		\For{every interval time $T_{\mathrm{update}}$}
			\State calculate $\delta=\mu\cdot(1-\mu)\cdot(q_{\mathrm{target}}-q_{\mathrm{avg}})$
			\If{$q_{\mathrm{avg}} < q_{\mathrm{target}}$}
				\State increase virtual target queue length: $\tilde{q}_{\mathrm{target}}=\min [q_{\max}-1\,,\,\tilde{q}_{\mathrm{target}}+\mathrm{Alpha}\cdot\delta ]$

			\ElsIf{$q_{\mathrm{avg}} > q_{\mathrm{target}}$}
				\State decrease virtual target queue length: $\tilde{q}_{\mathrm{target}}=\max [q_{\min}+1\,,\,\tilde{q}_{\mathrm{target}}+\mathrm{Beta}\cdot\delta ]$
			\EndIf
		\State update $\displaystyle \mu=\frac{\tilde{q}_{\mathrm{target}}-q_{\min}}{q_{\max}-q_{\min}}; \quad \sigma=\theta\cdot\sqrt{\mu(1-\mu)};$
		\EndFor 
		\State update $p=p_{\max } \cdot  \tilde{I}_{z(q_{\mathrm{avg}})}(\mu,\sigma)$
		\State with probability $p$, \textbf{drop arriving packet}
	\EndFor
\end{algorithmic}
\end{algorithm}

In this case we make a straightforward selection of the following parameters:
\begin{equation*}
	q_{\min}=0,  \quad  q_{\max}=B, \quad p_{\max}=1.
\end{equation*}
The choice of the minimum and maximum thresholds is based on the same reason as we have stated for the ABetaRED algorithm. Regarding $p_{\max}$, it is well known that the choice of $p_{\max}$ in RED is linked to the traffic load in the network, where the higher the traffic load the higher the value of $p_{\max}$ should be. One of the advantages of BetaRED over RED is that we can control the increase of the drop probability function around the target queue length (in a smoothly or sharply way, but always continuously) by means of the $\theta$ parameter. Thus, the choice of $p_{\max}=1$ is the most reasonable in order to avoid discontinuities in the drop probability function. Moreover, the numerical results of Figure \ref{fig-beta-red-pmax-N} show that a higher value of $p_{\max}$ implies better performance. The control parameters $\mathrm{Alpha}$ and $\mathrm{Beta}$  determine the size of the increase and decrease steps of the virtual target queue length, respectively, with the range of admissible values being $0<\mathrm{Alpha},\mathrm{Beta}\leq 1$. The tunable parameters to be set by the user in this case are the target delay $T_{\mathrm{target}}$, the scalar factor $\theta$ and the averaging weight $w$.

\subsection{Simulations} \label{subsec-dynamic-beta-red-simulations}

This section presents the results and discussion of the numerical simulations performed to compare ABetaRED (Section \ref{subsec-adaptative-beta-red}) and DBetaRED (Section \ref{subsec-dynamic-beta-red}), as well as these algorithms with the selection of the other AQM algorithms described in Section \ref{sec-related-works}. The simulations will be performed in several scenarios according to different parameters and levels of congestion, including slight, moderate and abrupt variations in the number of nodes in order to verify the robustness of the proposed AQM algorithms. However, space limitation conditions us to present only a selection of the most important features regarding the behavior of the above 
algorithms.

\begin{figure}[htbp]
	\centering
	\includegraphics{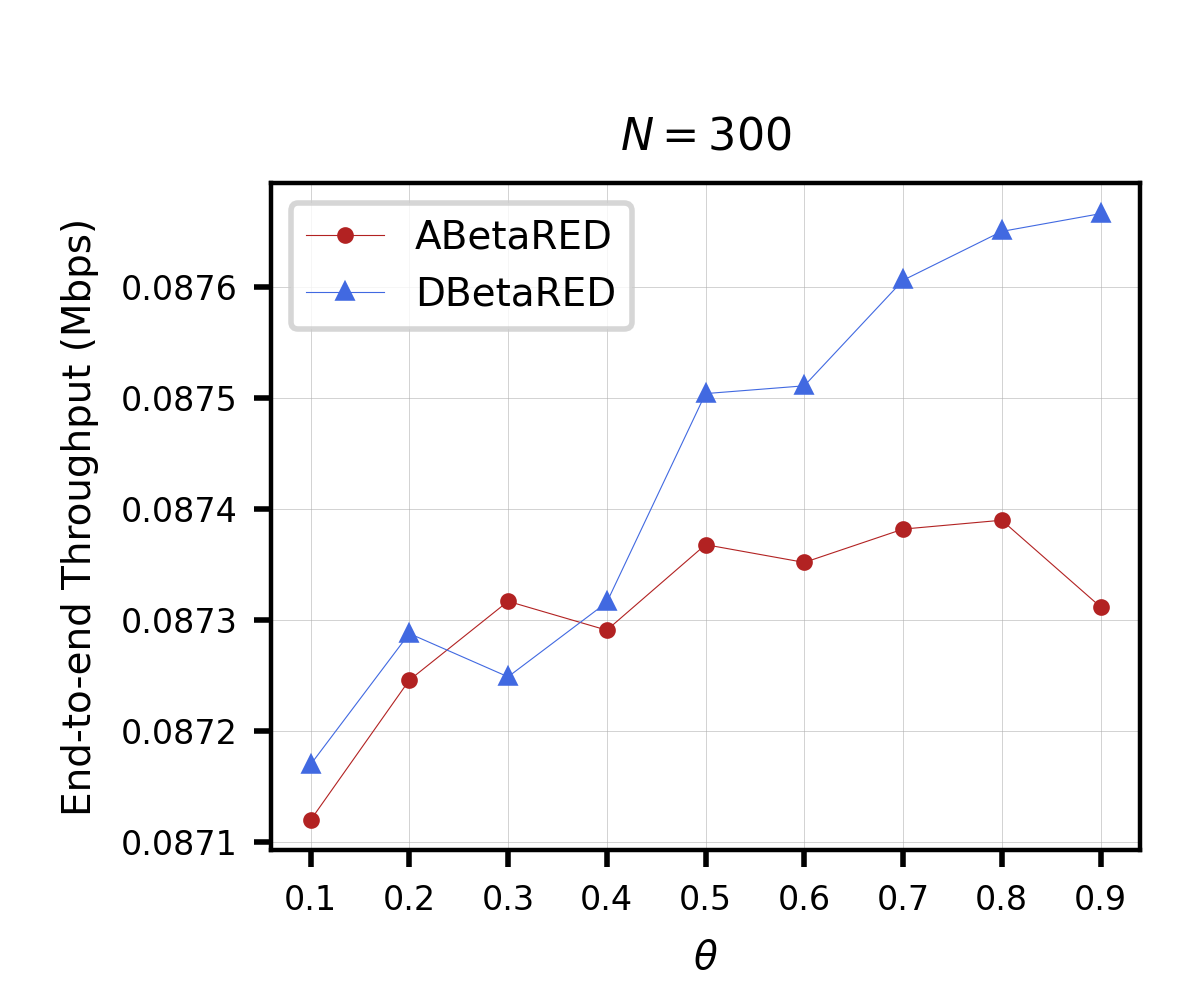}
	\includegraphics{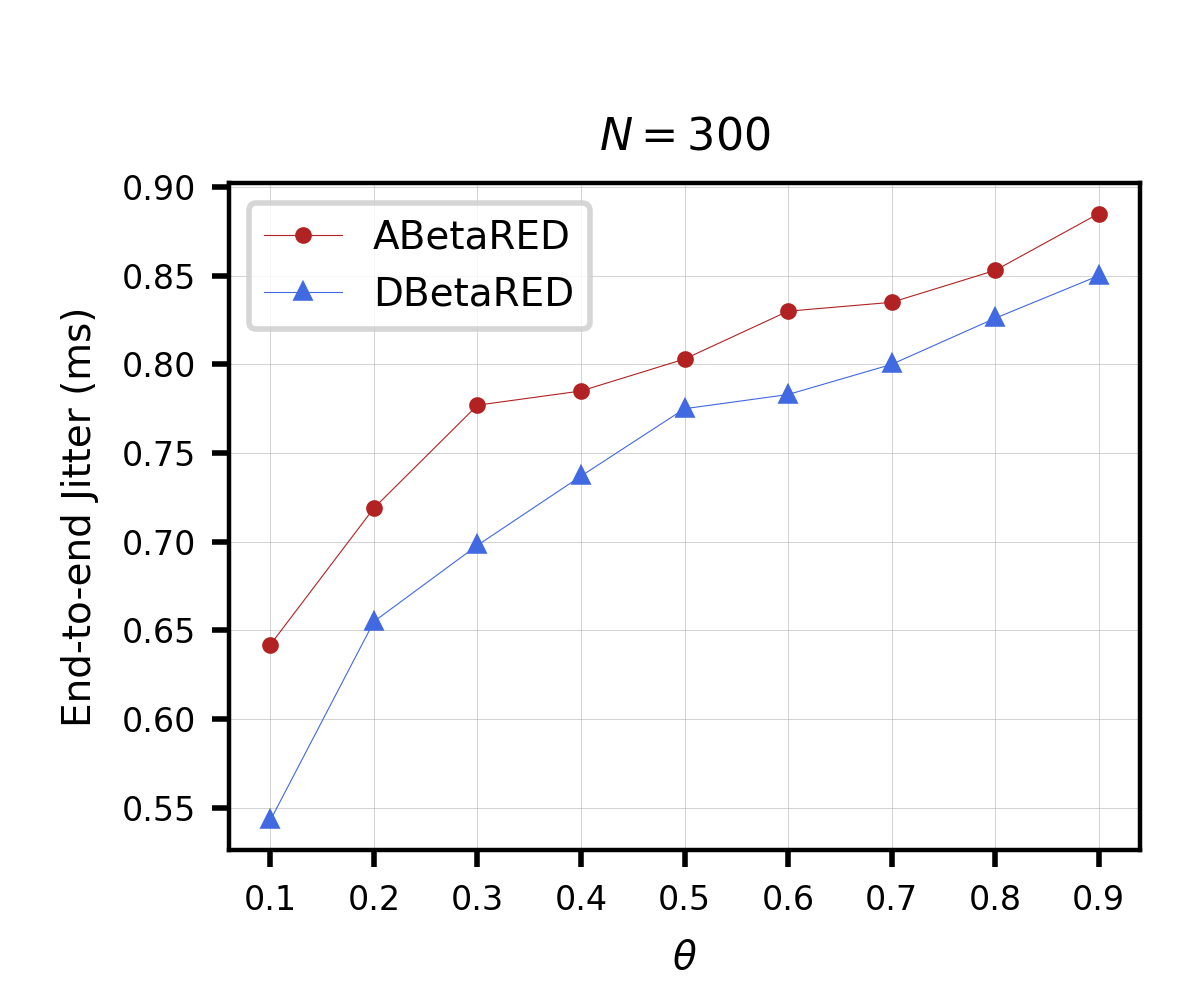}
	\includegraphics{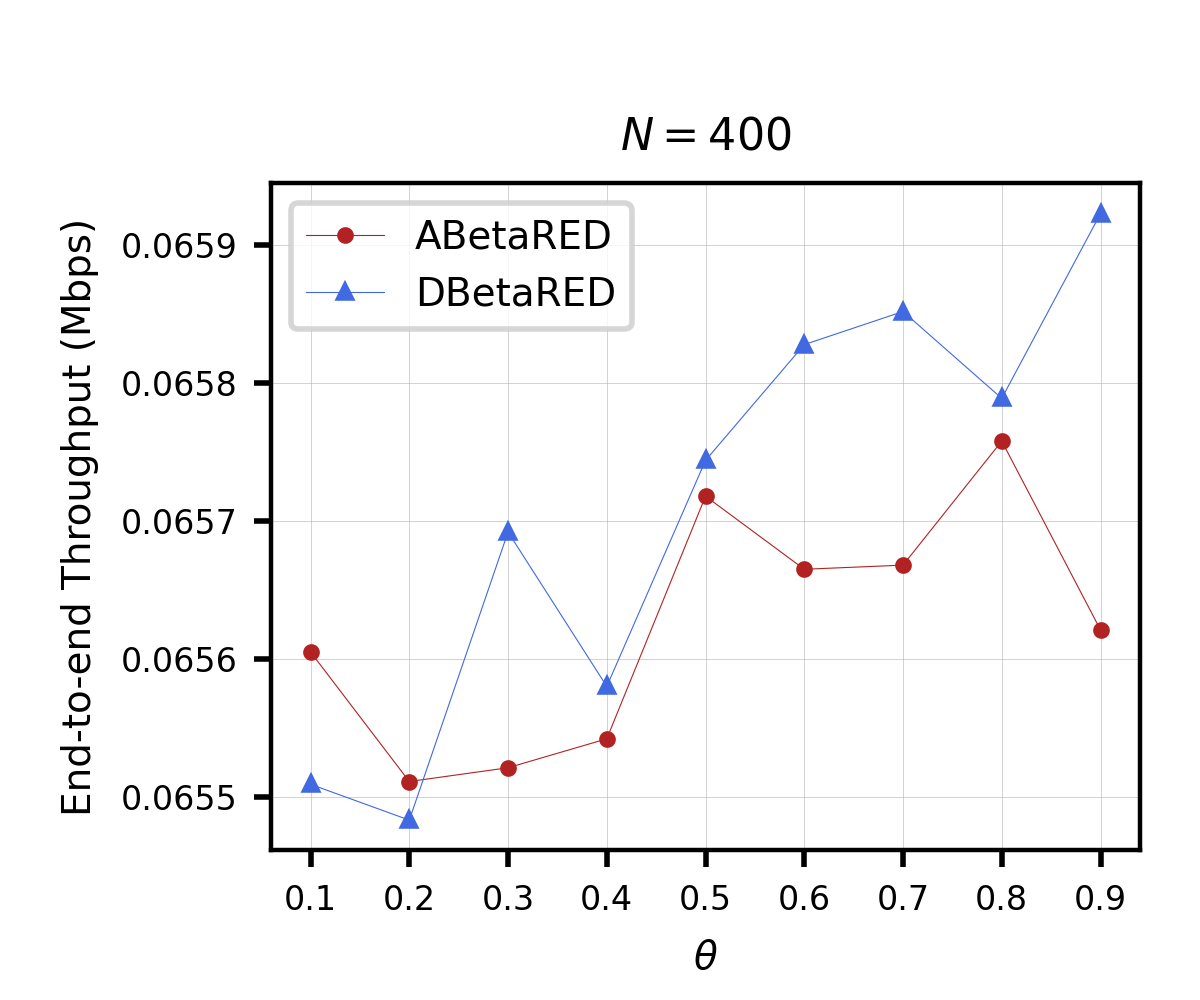}
	\includegraphics{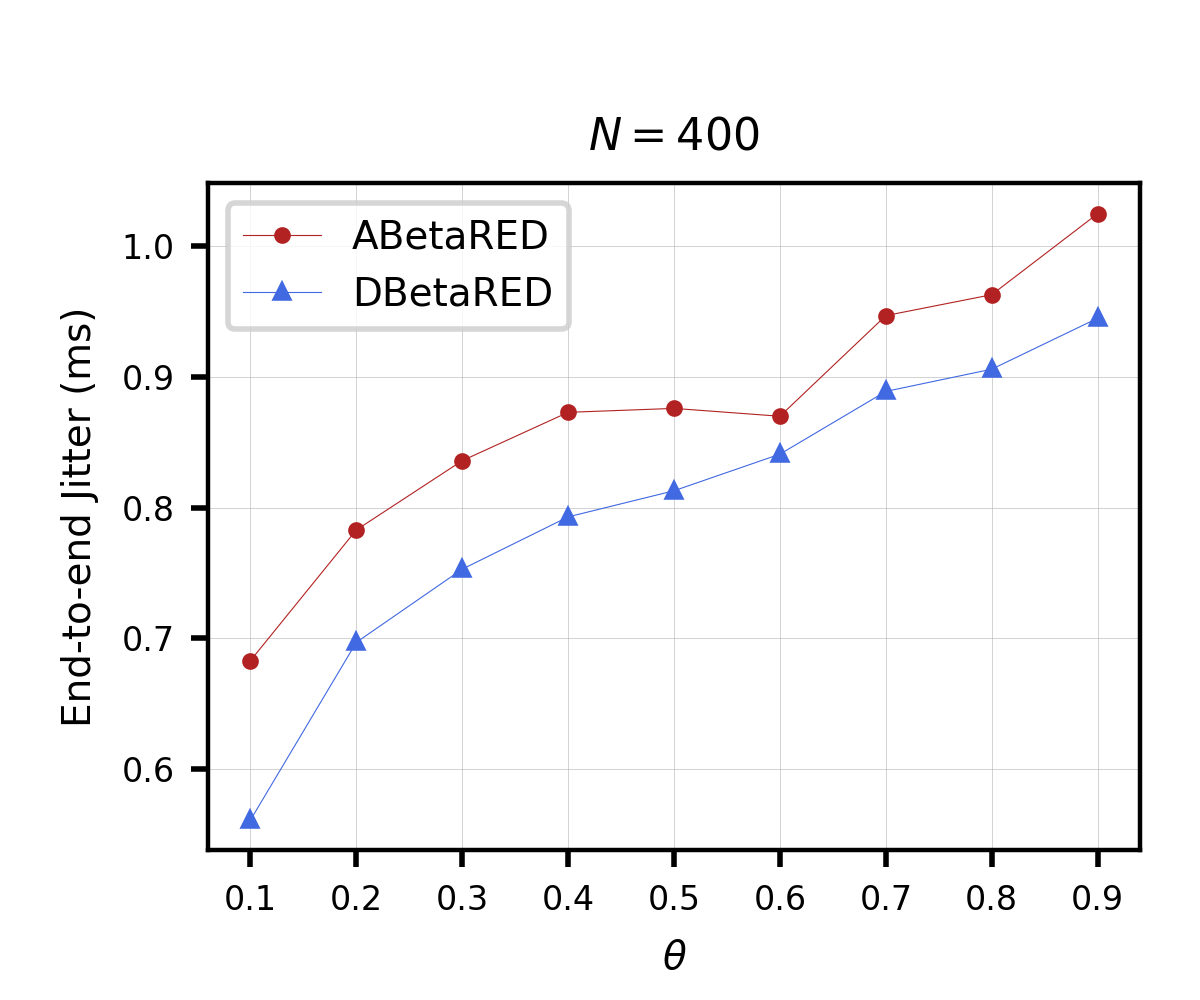}
	\includegraphics{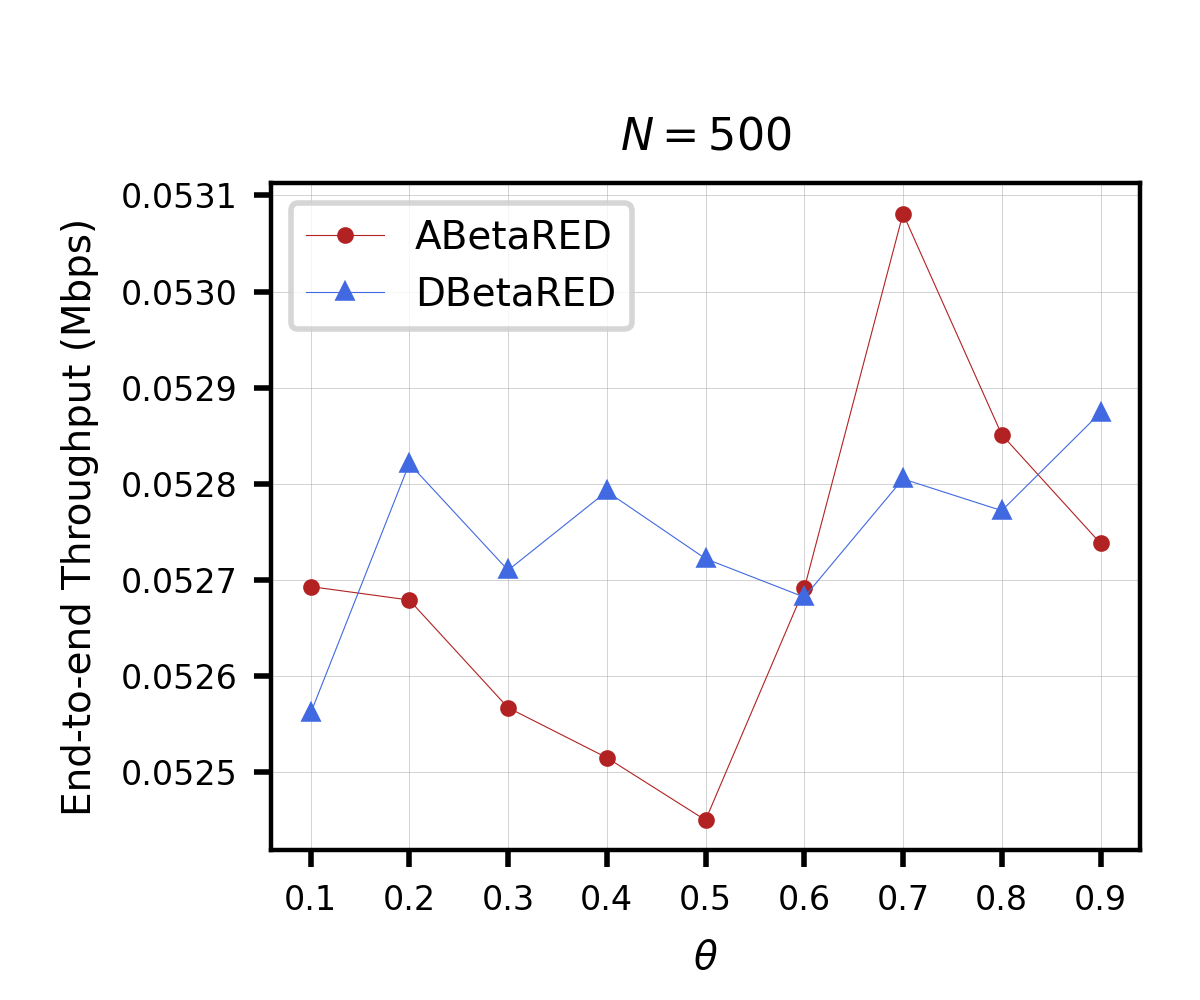}
	\includegraphics{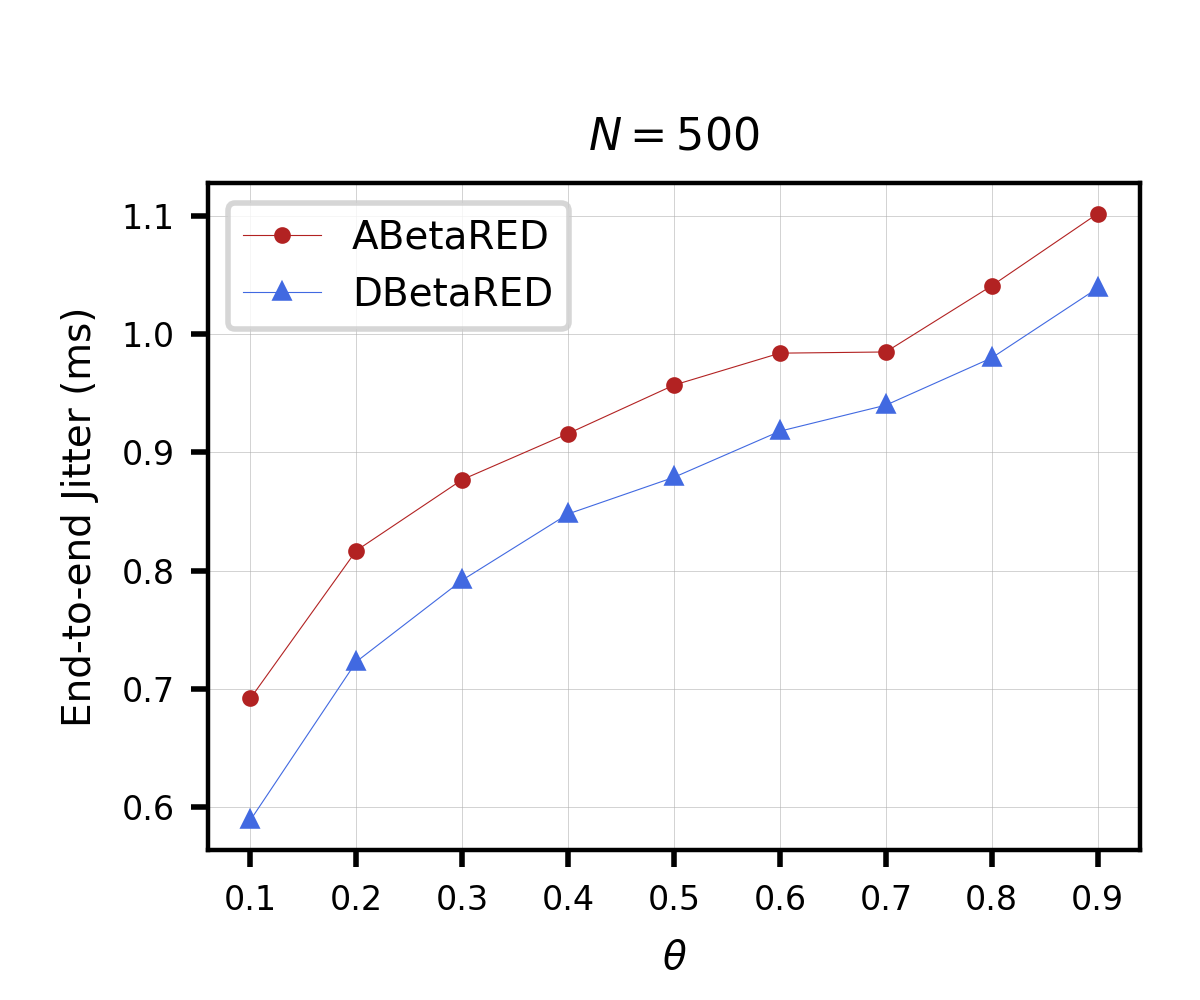}

		\caption{Comparison between ABetaRED and DBetaRED algorithms according to the standard deviation $\sigma$. Different levels of congestion are also considered by means of a constant number $N$ of active flows. Other tuning parameters:  $T_{\mathrm{target}}=40\,\mathrm{ms}$ ($q_{\mathrm{target}}=250\,\mathrm{packets}$) and $w=0.1$. Scenario 1 on the dumbbell topology (described in Section \ref{subsec-scenarios-simulation}) is used. Overall, the DBetaRED algorithm exhibits better performance, especially for the end-to-end jitter metric.}
	
	\label{fig-sigma-abetared-vs-dbetared}
\end{figure}

\begin{figure}[htbp]
	\centering
	\includegraphics{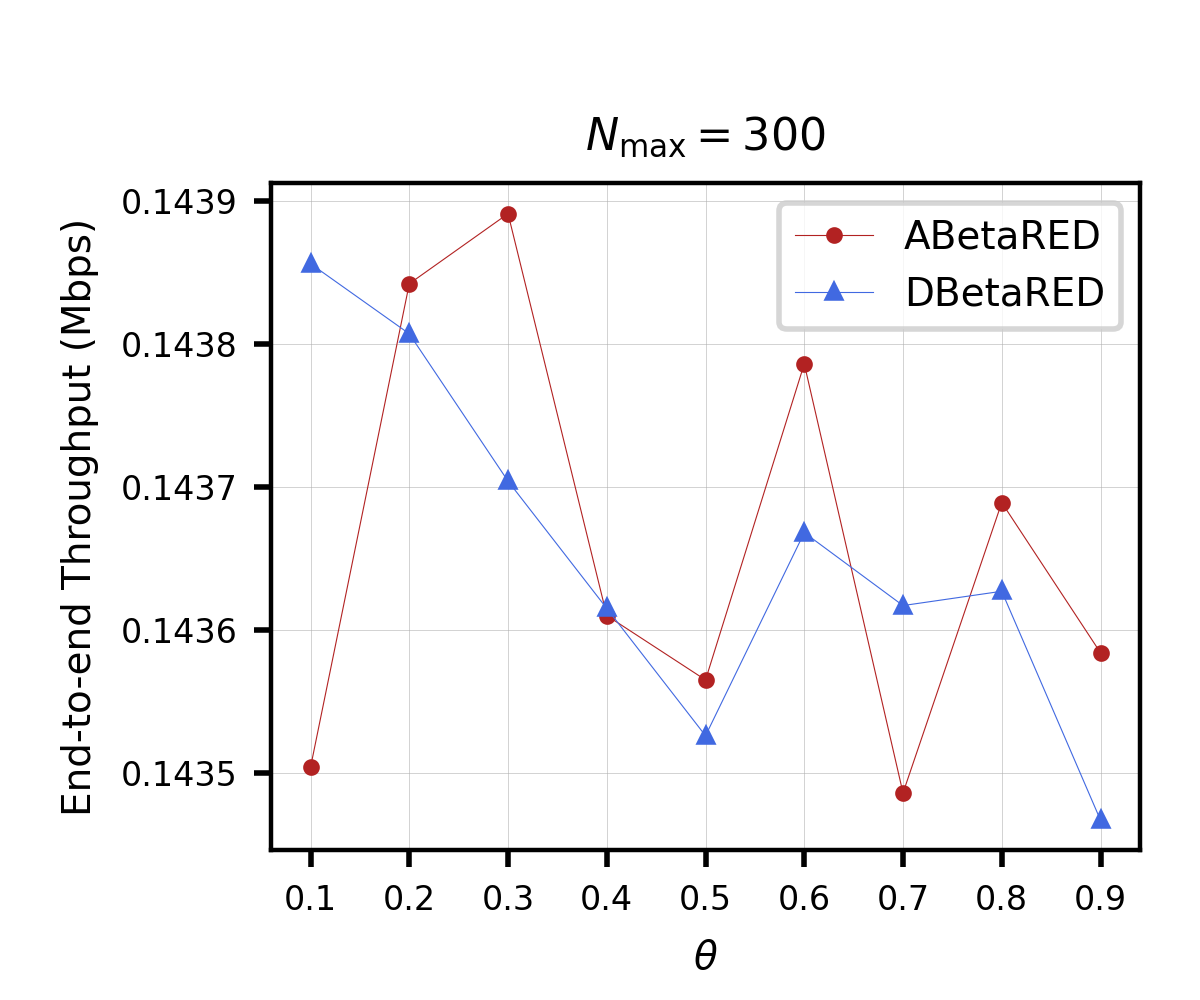}
	\includegraphics{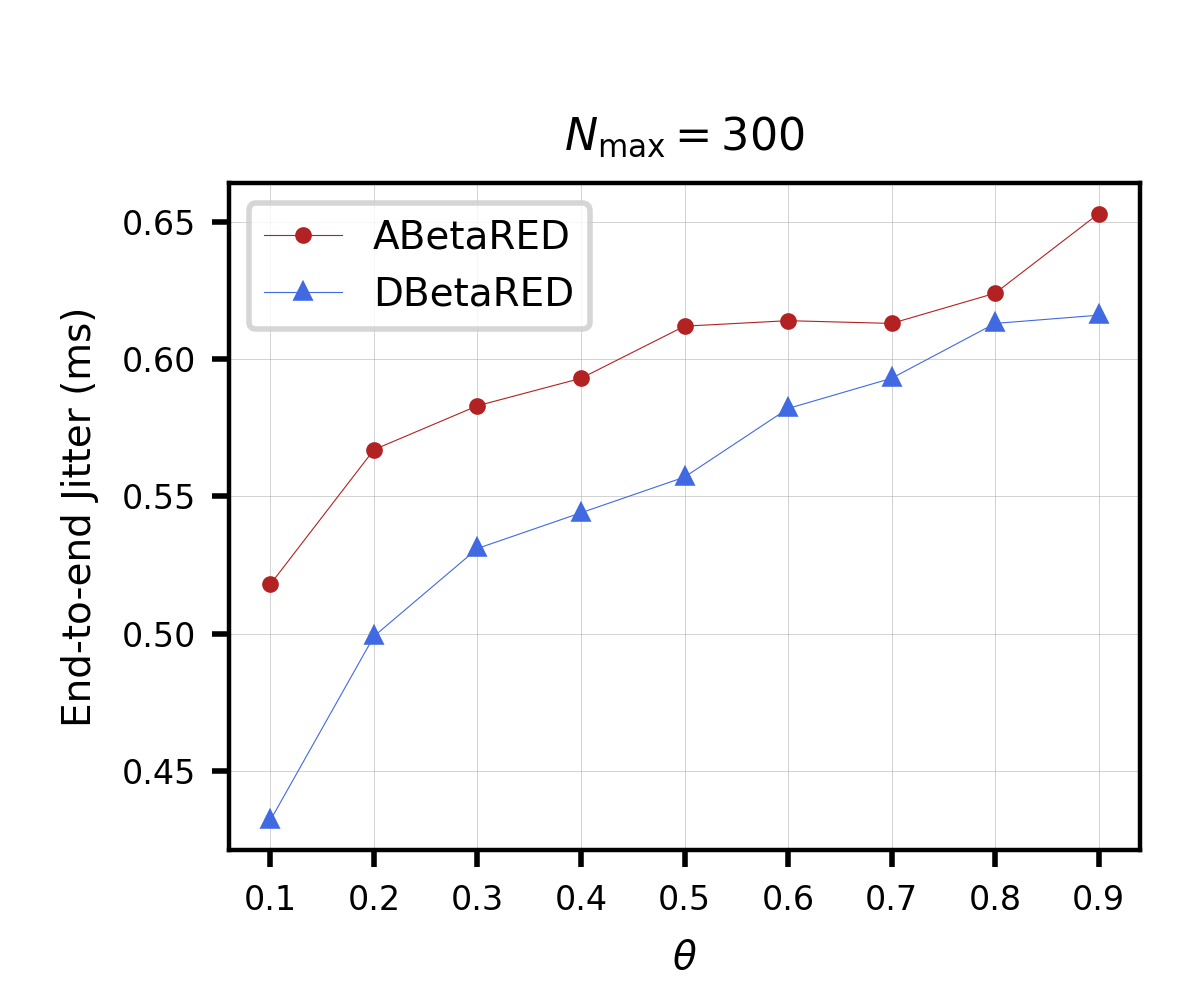}
	\includegraphics{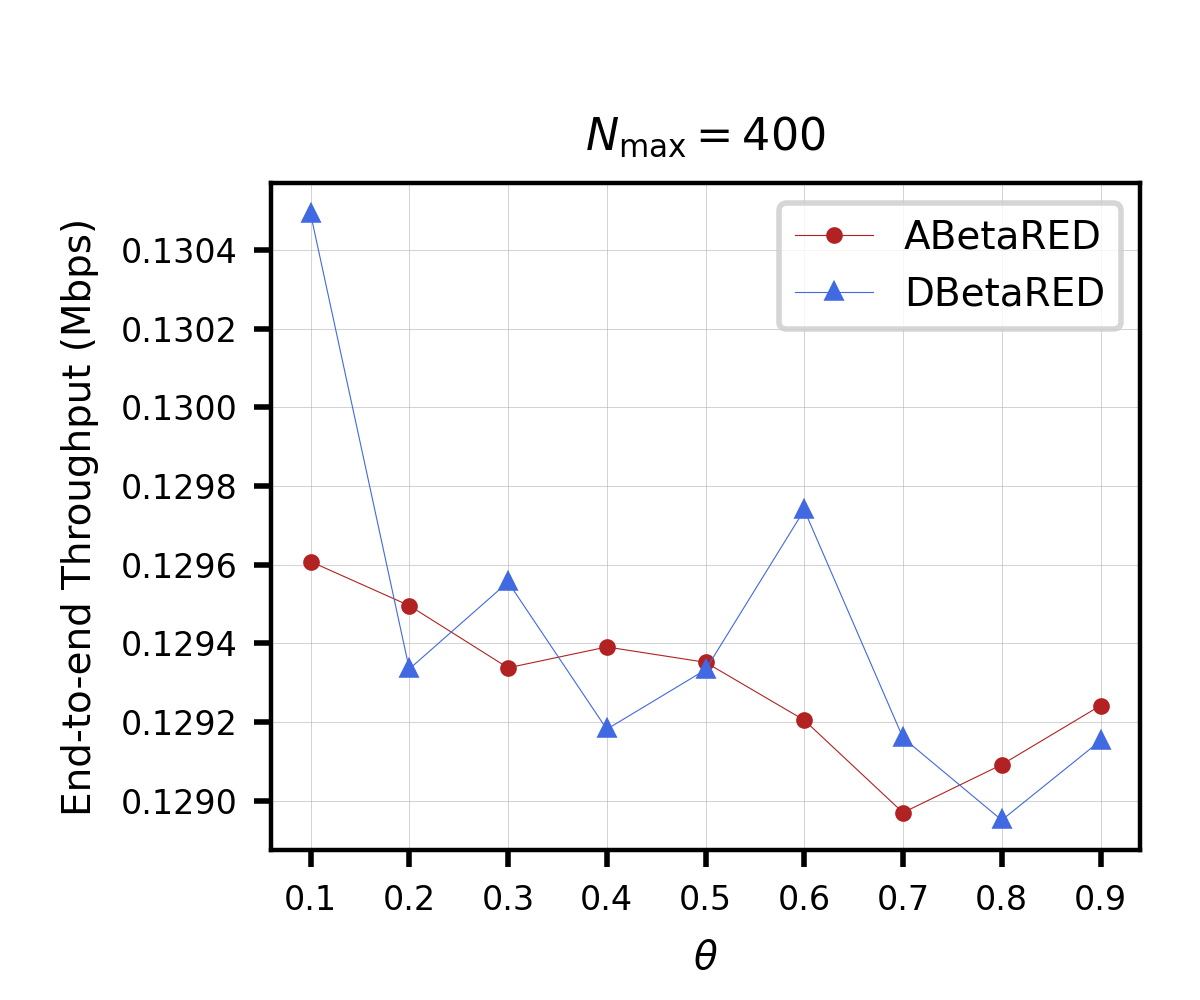}
	\includegraphics{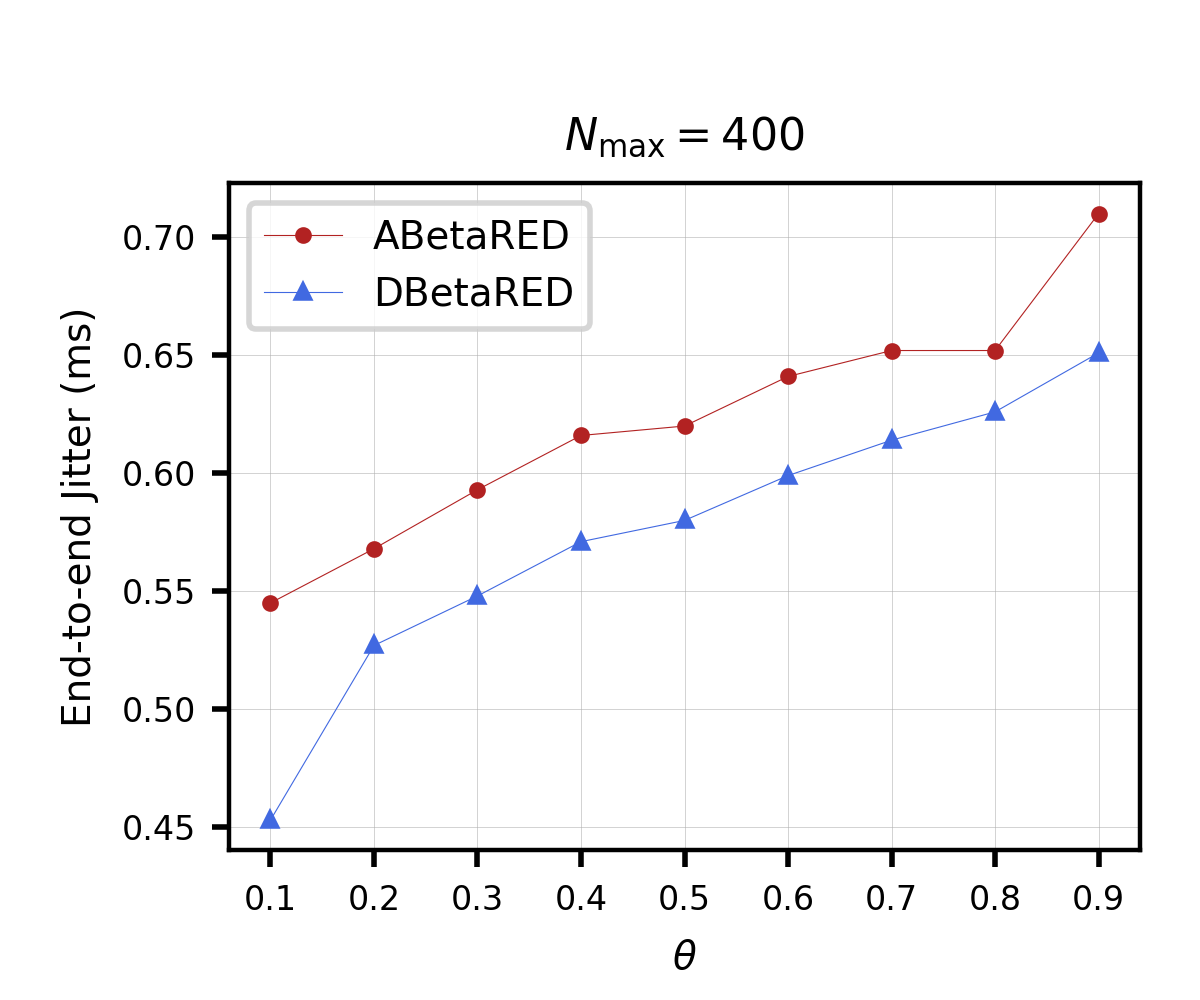}
	\includegraphics{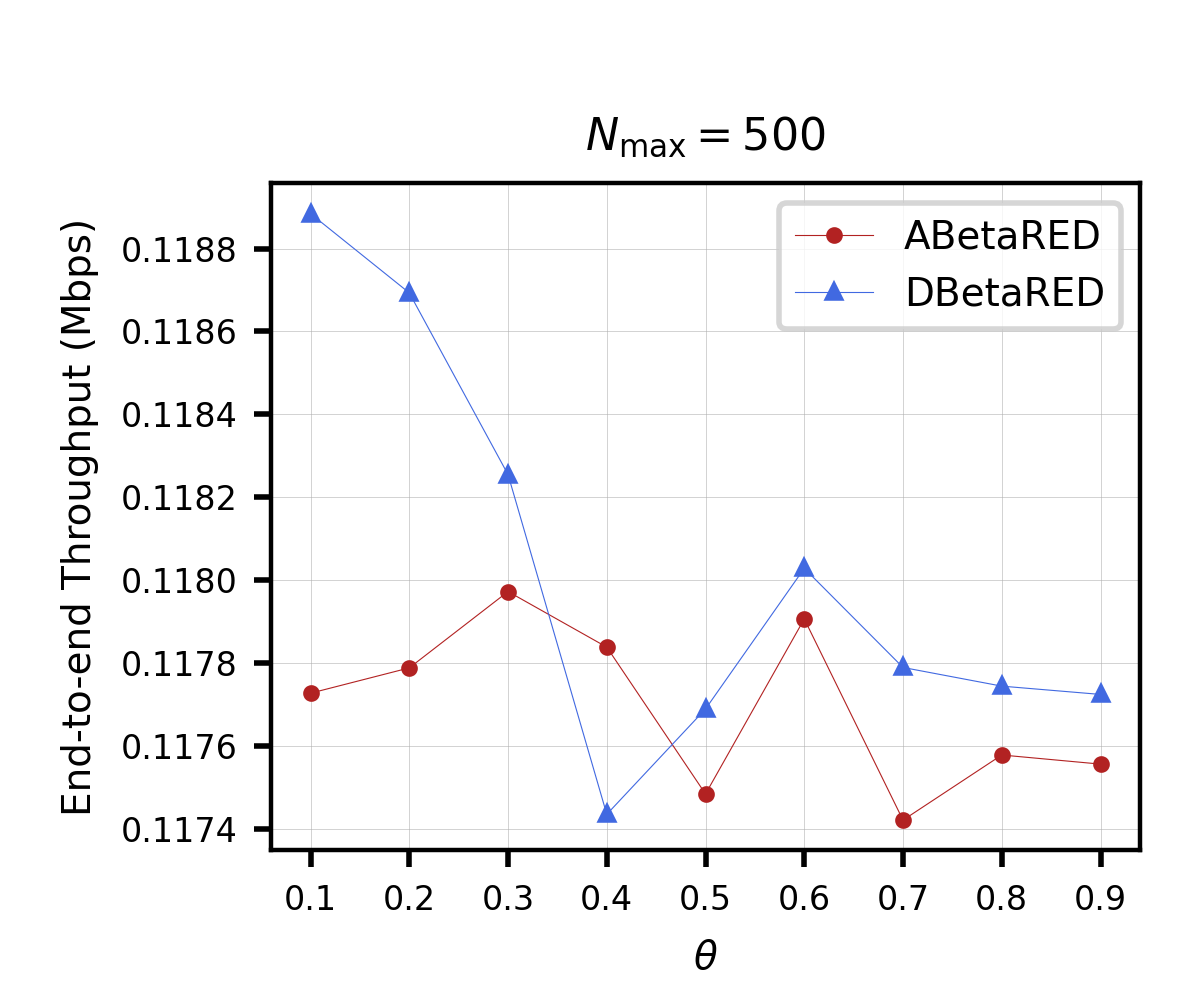}
	\includegraphics{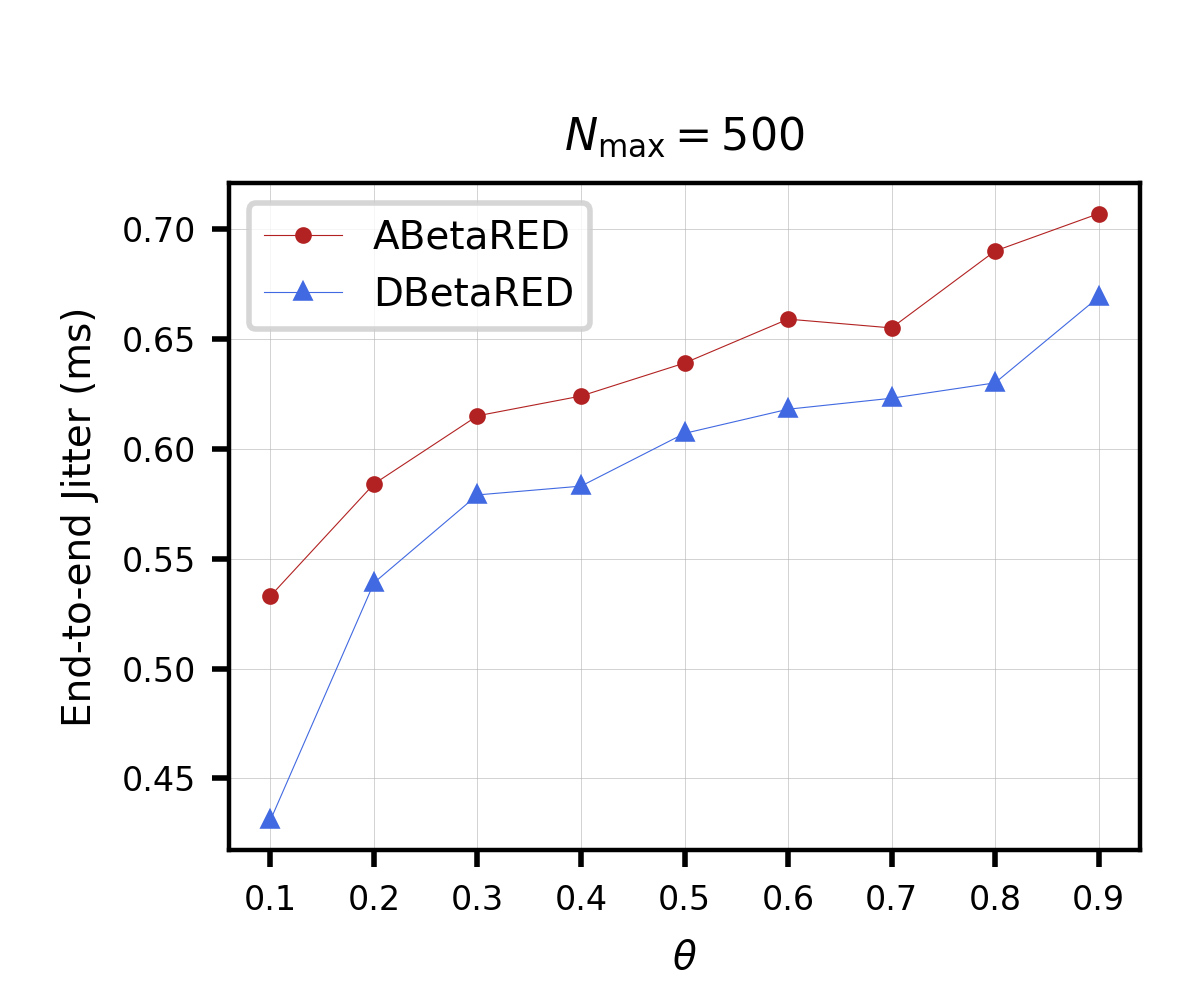}

	\caption{Comparison between the ABetaRED and DBetaRED algorithms according to the parameter $\theta$. Different levels of congestion are also considered, but, unlike in Figure \ref{fig-sigma-abetared-vs-dbetared}, the number $N$ of active flows varies dynamically with time. Other tuning parameters:  $T_{\mathrm{target}}=40\,\mathrm{ms}$ ($q_{\mathrm{target}}=250\,\mathrm{packets}$) and $w=0.1$. Scenario 2 on the dumbbell topology (described in Section \ref{subsec-scenarios-simulation}) is used. As in the non-dynamic scenario (Figure \ref{fig-sigma-abetared-vs-dbetared}), the DBetaRED algorithm performs better than the ABetaRED algorithm.}
	
	\label{fig-sigma-abetared-vs-dbetared-dynamic}
\end{figure}

The first comparison we carried out was between the ABetaRED and DBetaRED algorithms (see Figures \ref{fig-sigma-abetared-vs-dbetared} and \ref{fig-sigma-abetared-vs-dbetared-dynamic}) when we progressively increase the parameter $\theta$ (and hence the standard deviation $\sigma$). Figure \ref{fig-sigma-abetared-vs-dbetared} shows the result of the simulations for a constant level of congestion (Scenario 1 described in Section \ref{subsec-scenarios-simulation}), while Figure \ref{fig-sigma-abetared-vs-dbetared-dynamic} shows the result for a changing level of congestion (Scenario 2 described in Section \ref{subsec-scenarios-simulation}). It is observed that, in both algorithms, the throughput is quite unpredictable for different $\theta$ parameters. However, the jitter is lower the smaller the value of the $\theta$ parameter, i.e., for smaller $\sigma$ values.

\begin{figure}[htbp]
	\centering
	\includegraphics{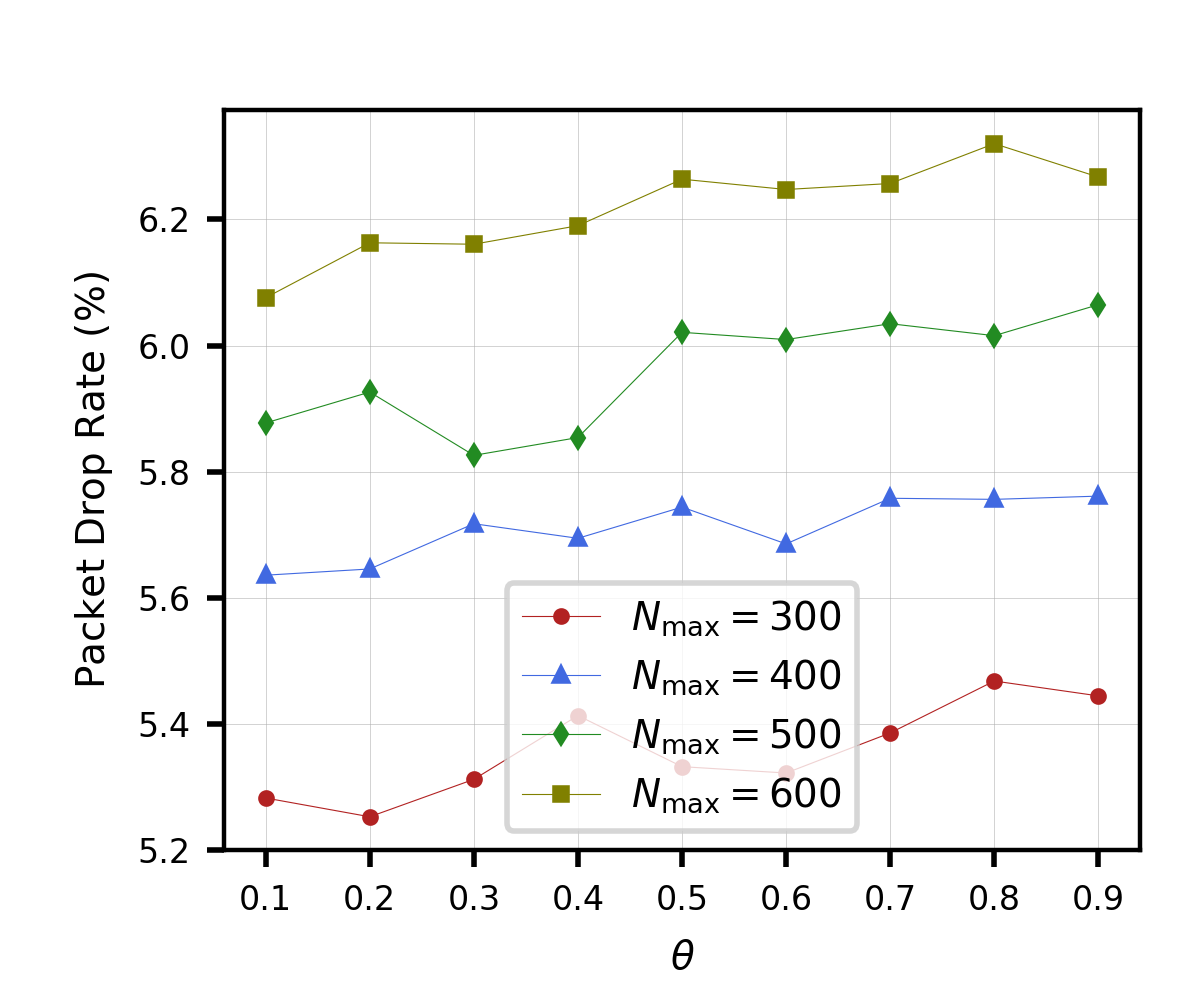}
	\includegraphics{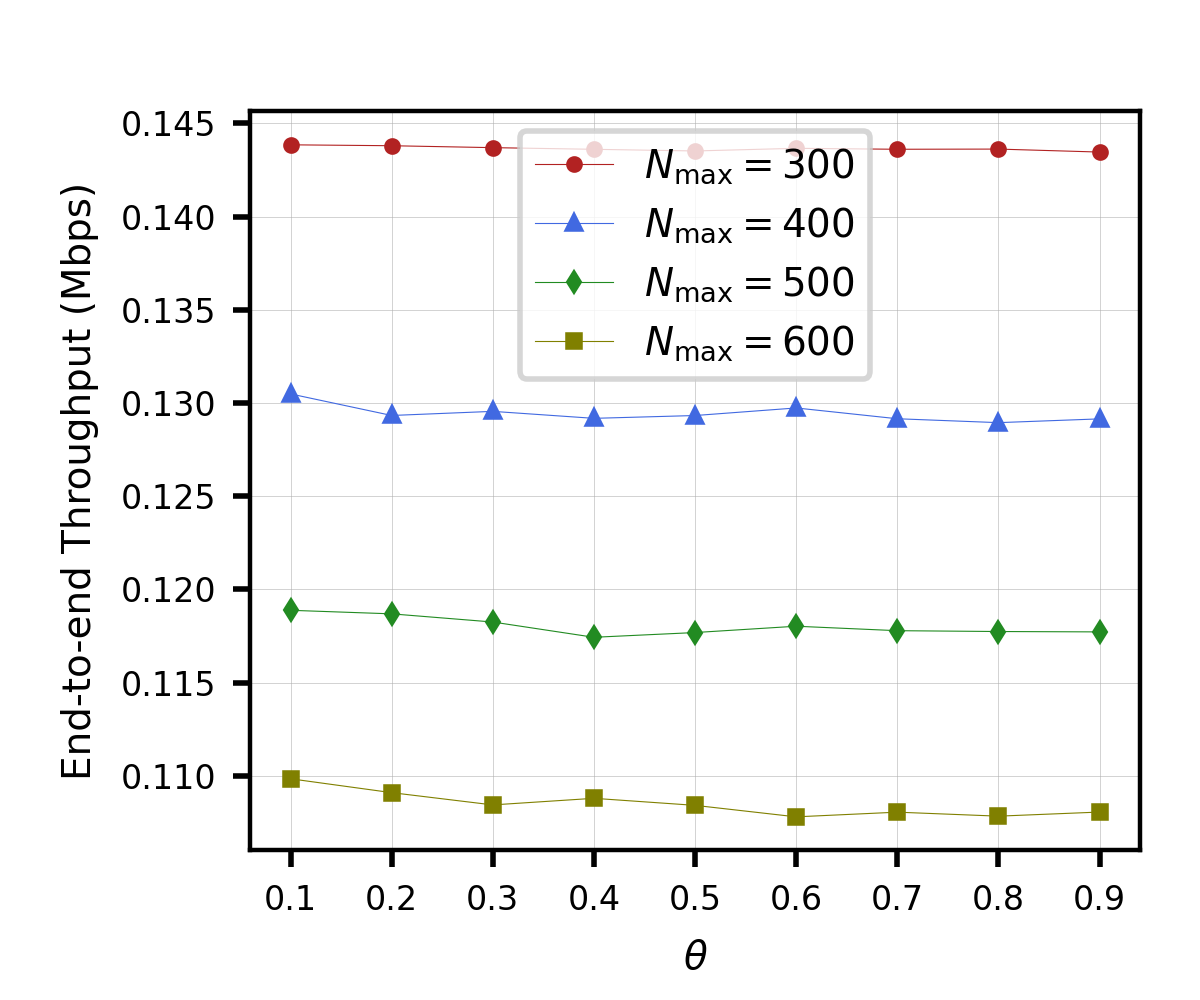}
	\includegraphics{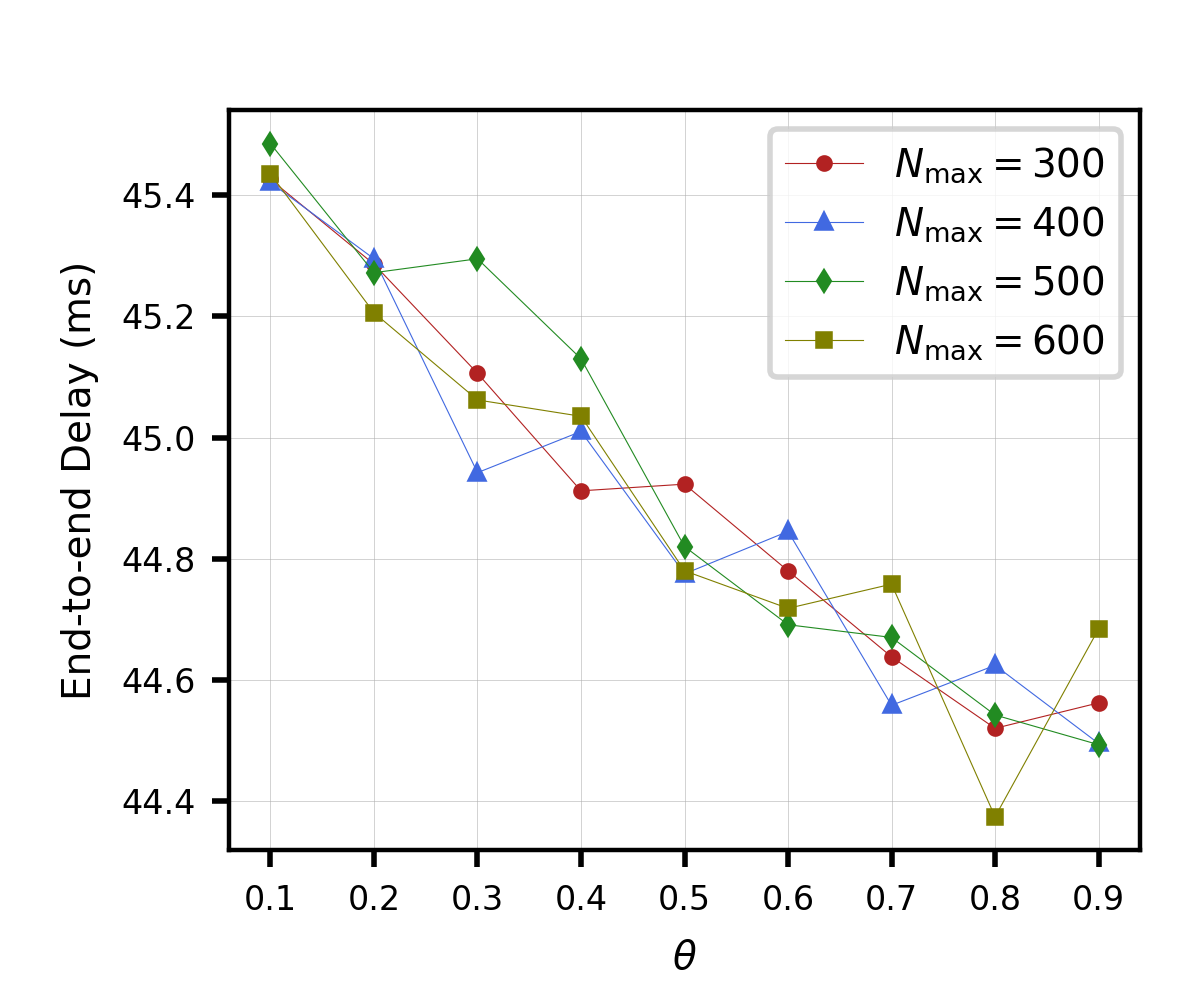}
	\includegraphics{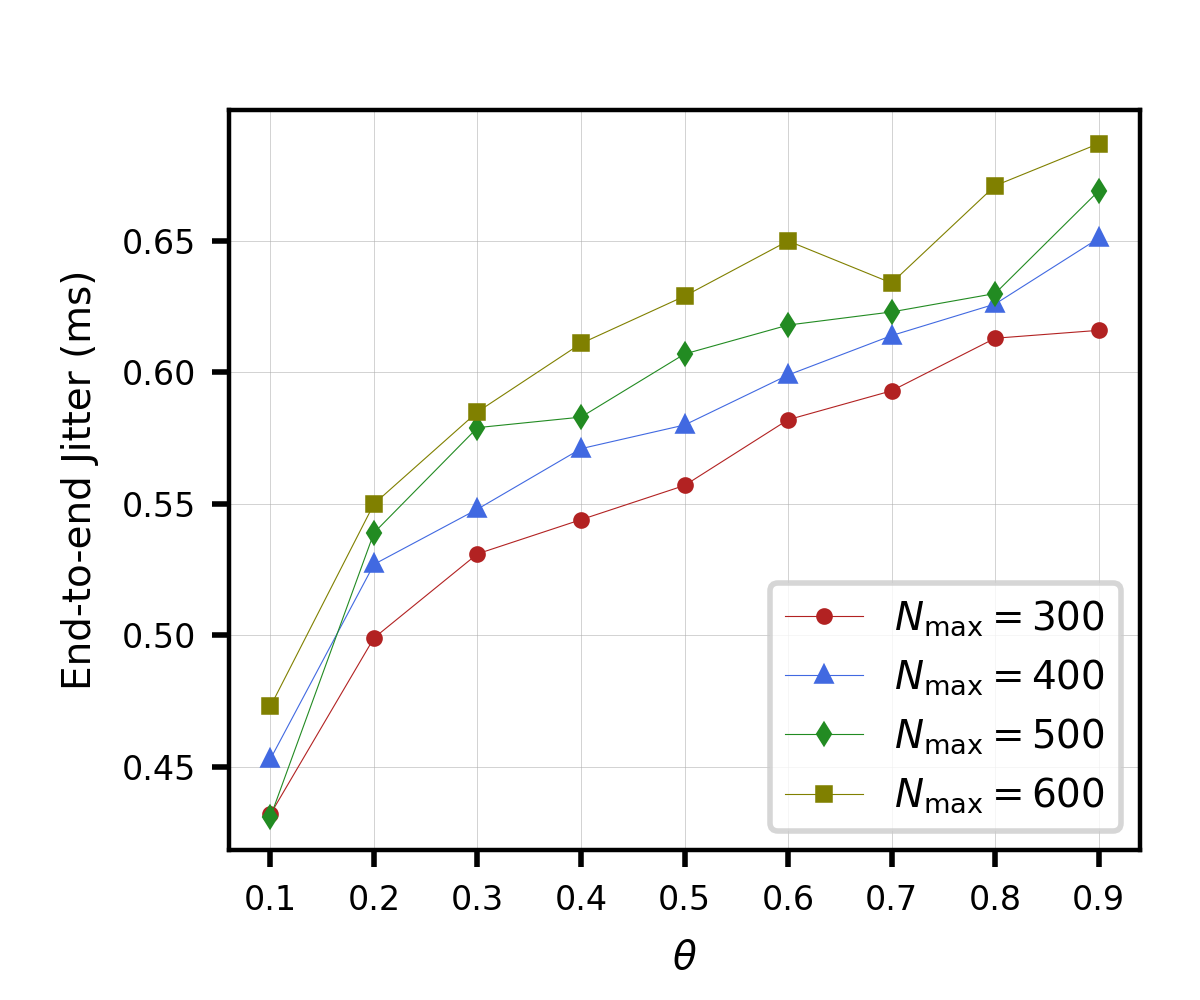}

	\caption{Performance comparison of the Dynamic Beta RED algorithm according to a number $N$ of active flows that vary dynamically over time and different values of the weight parameter $w$. Other tunable parameters:  $T_{\mathrm{target}}=40\,\mathrm{ms}$ ($q_{\mathrm{target}}=250\,\mathrm{packets}$) and $\theta=0.1$. Scenario 2 on the dumbbell topology (described in Section \ref{subsec-scenarios-simulation}) is used. We observe that the trend is for the performance of the DBetaRED algorithm to increase as the value of $w$ increases.}
	
	\label{fig-dbetared-sigma-weight}
\end{figure}

Figure \ref{fig-dbetared-sigma-weight} shows the performance of DBetaRED for different levels of congestion according to different values of the weight parameter $w$. It can be seen that whatever the level of congestion, the choice of a sufficiently high $w$ parameter guarantees a satisfactory performance. However, the most appropriate value of $w$ according to the scenario is difficult to estimate, as it may vary depending on the network topology, the level of congestion, etc. In any case, it was observed that the qualitative behavior does not change from one scenario to another.

It is of special interest when a comparison is made with other related AQM schemes. The common denominator of all the selected dynamic algorithms to be compared with ABetaRED and DBetaRED is the need to set the target delay parameter $T_{\mathrm{target}}$. However, the mechanism of action of each of the AQM algorithms according to this parameter is different. CoDel starts dropping packets when the queue delay remains above the target delay for a certain time. PIE continuously updates its probability of dropping packets according to the difference between the current queue delay and the target delay. ARED does not act directly as a function of the target delay, but rather as a function of a target queue length estimated via the target delay. Both ABetaRED and DBetaRED algorithms  act in the same way as ARED, namely, to achive a given predetermined $T_{\mathrm{target}}$, a target queue length given by $q_{\mathrm{target}}=C\cdot T_{\mathrm{target}}$ is estimated, where $C$ is measured in packets per second.

\begin{figure}[htbp]
	\centering
	\includegraphics{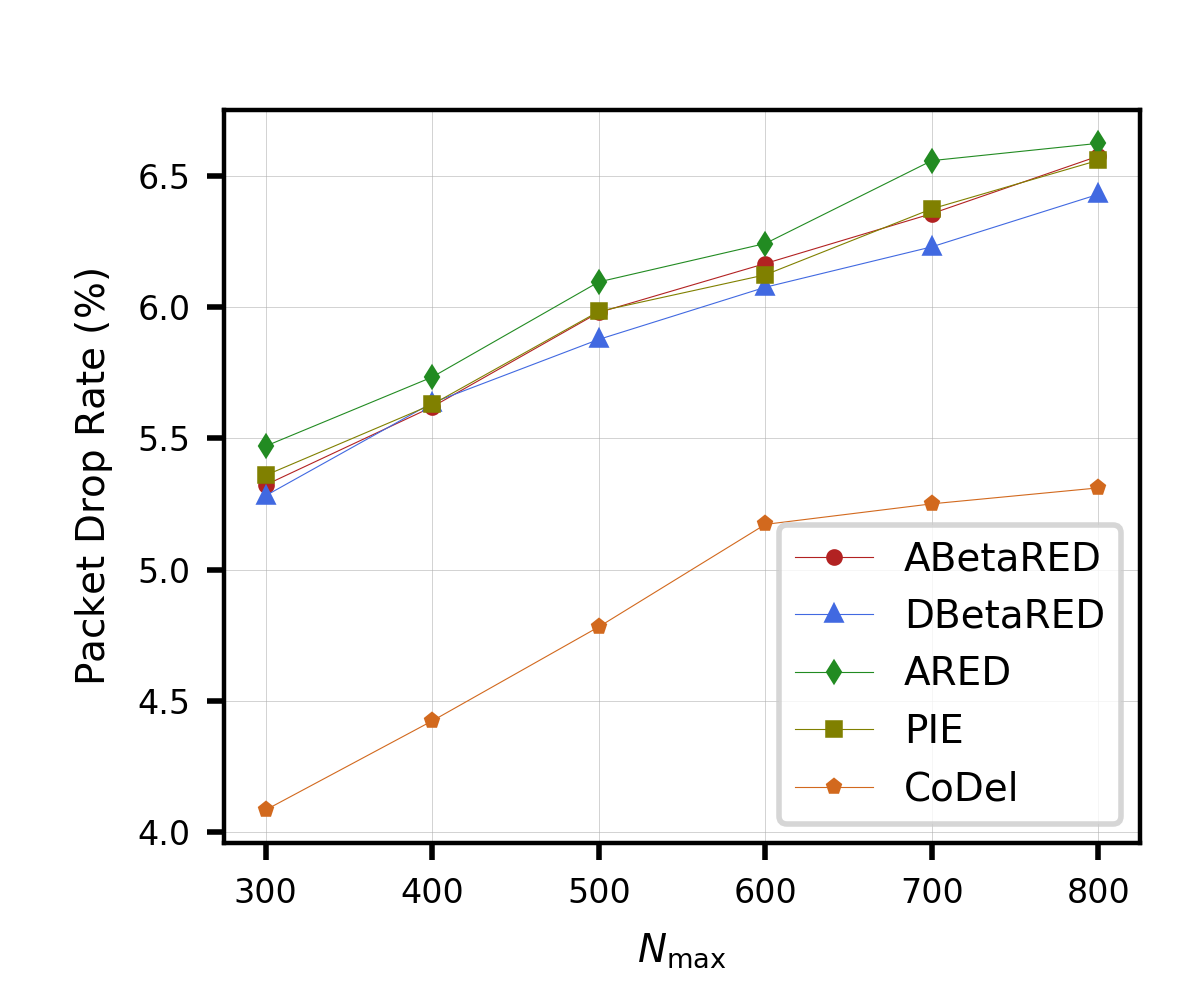}
	\includegraphics{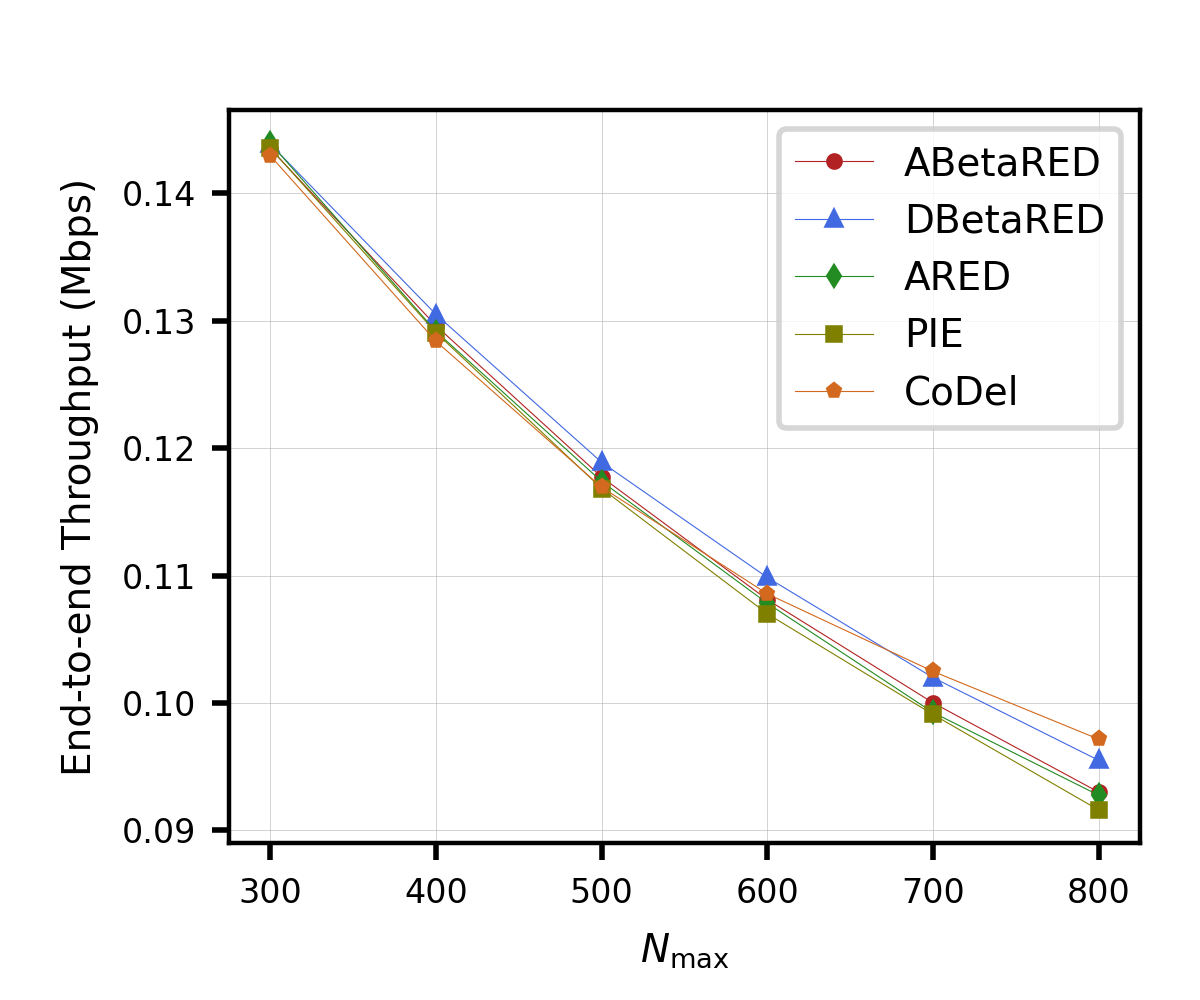}
	\includegraphics{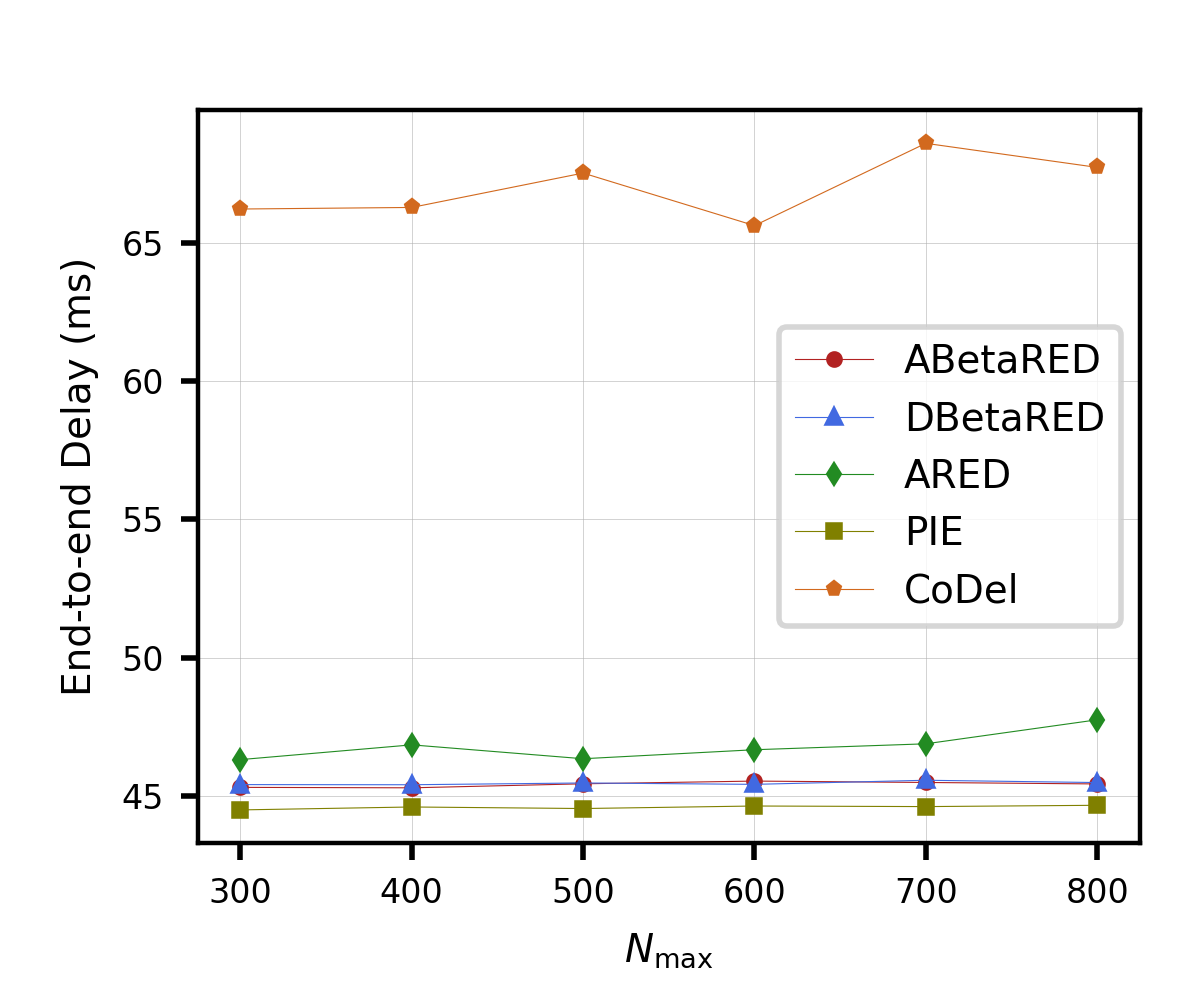}
	\includegraphics{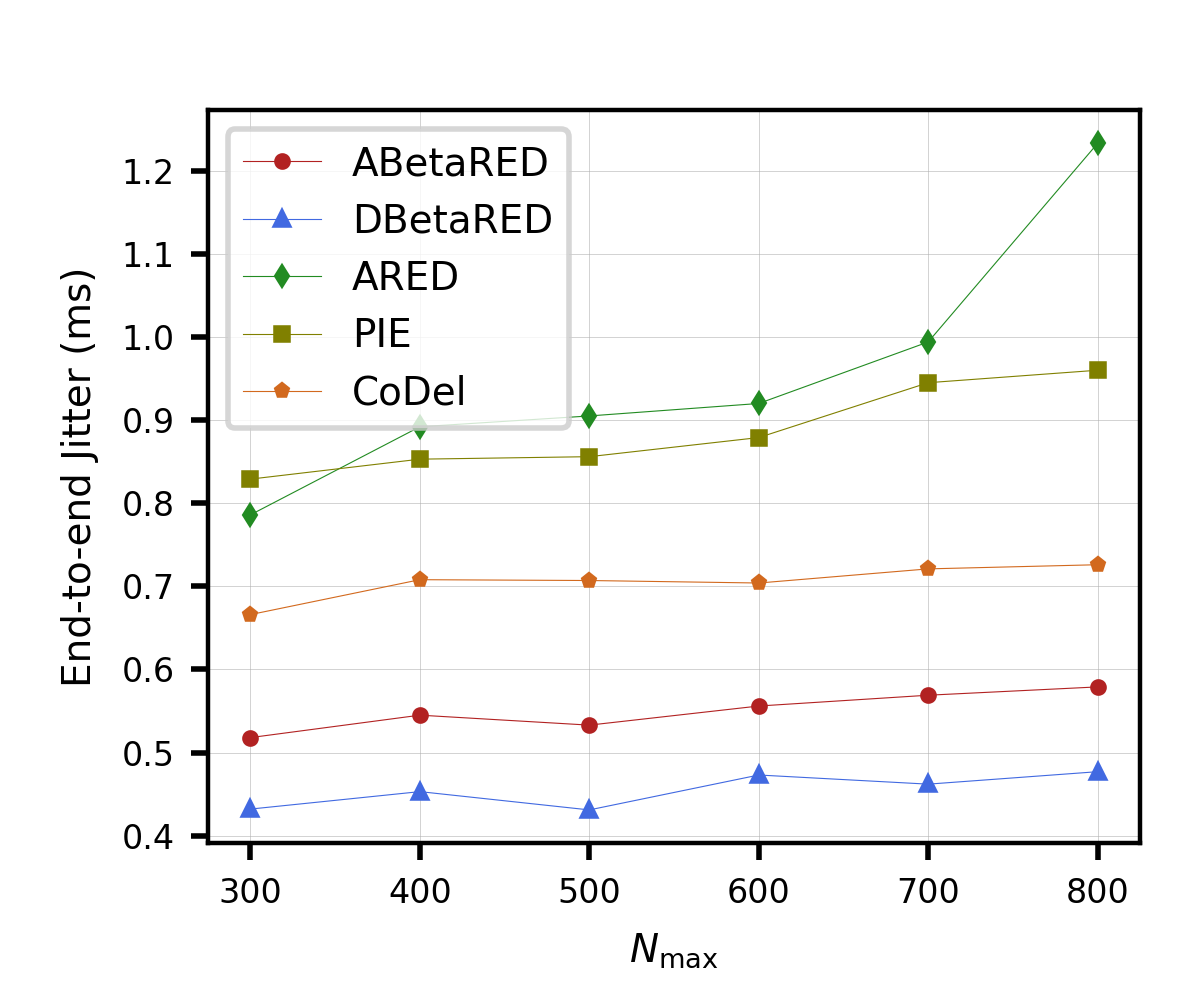}
	
	\caption{Performance comparison between related AQM algorithms when considering different levels of congestion by varying dynamically the number $N$ of active flows. Other tuning parameters:  $T_{\mathrm{target}}=40\,\mathrm{ms}$ for all AQM algorithms; $\theta=0.1$  and $w=0.1$ for ABetaRED and DBetaRED. Scenario 2 on the dumbbell topology (described in Section \ref{subsec-scenarios-simulation}) is used. The performance of the DBetaRED algorithm outperforms the other algorithms as the congestion level increases.}

	\label{fig-comparison-all-algorithms}
\end{figure}

\begin{figure}[htbp]
	\centering
	\includegraphics{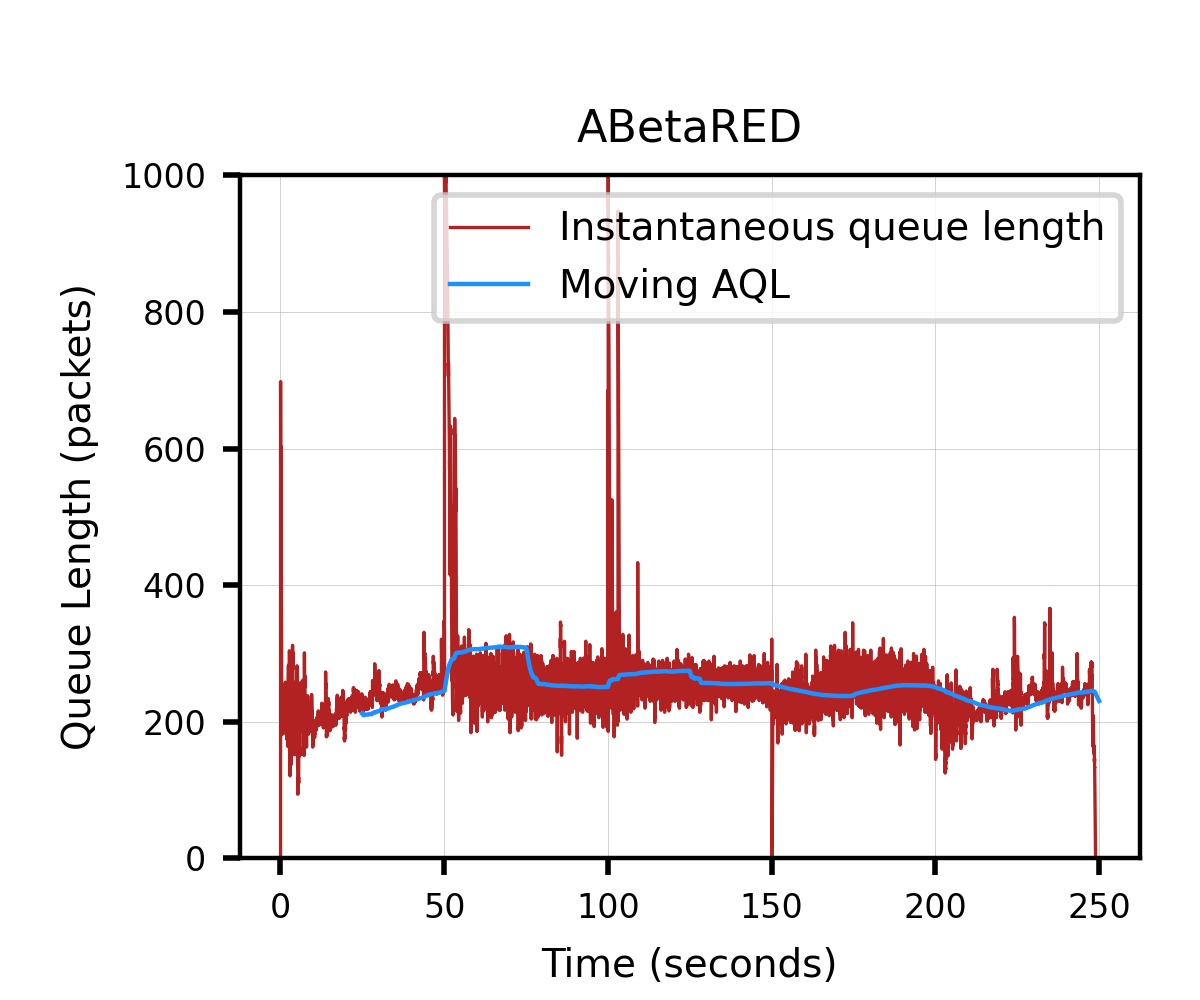}
	\includegraphics{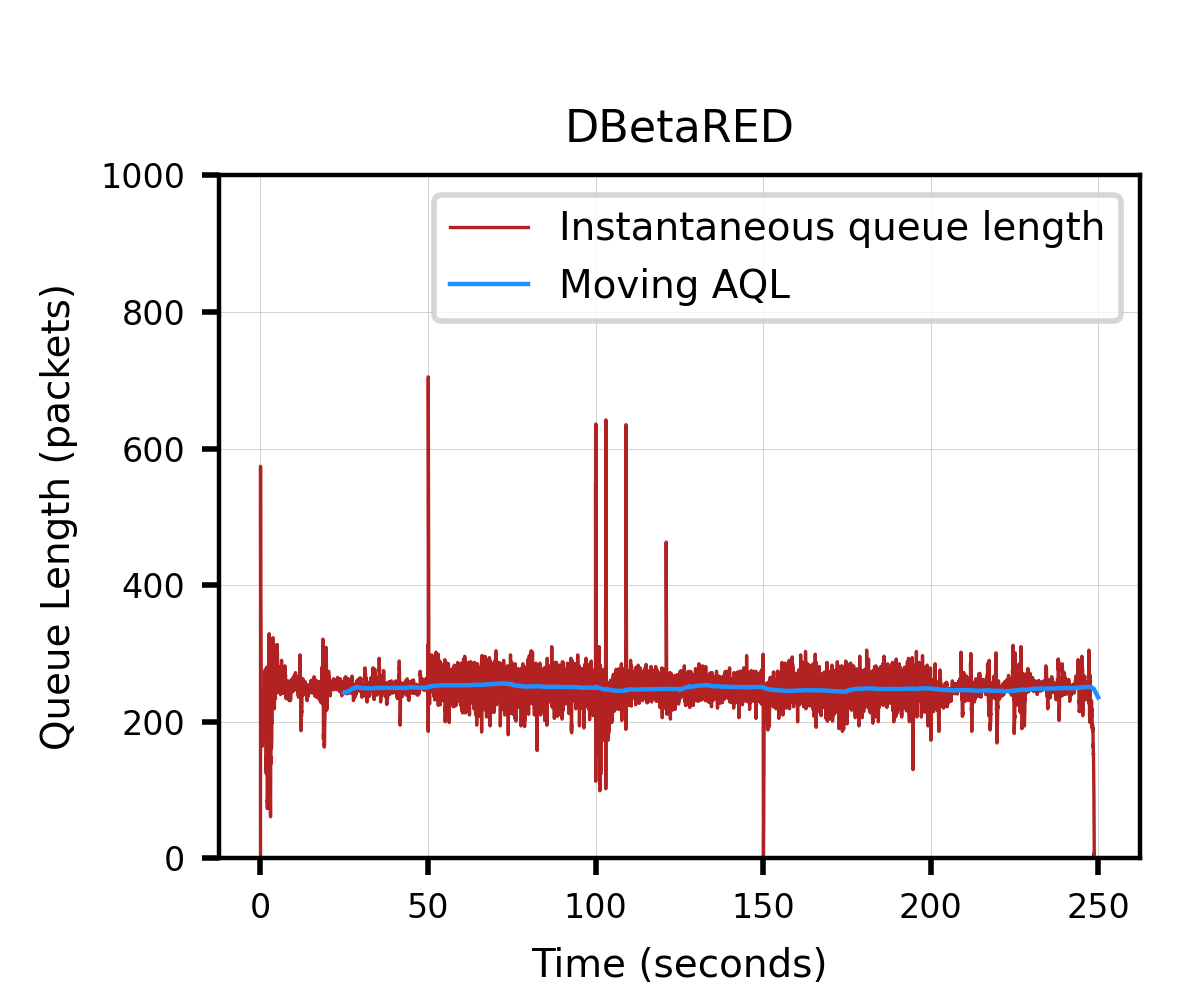}
	\includegraphics{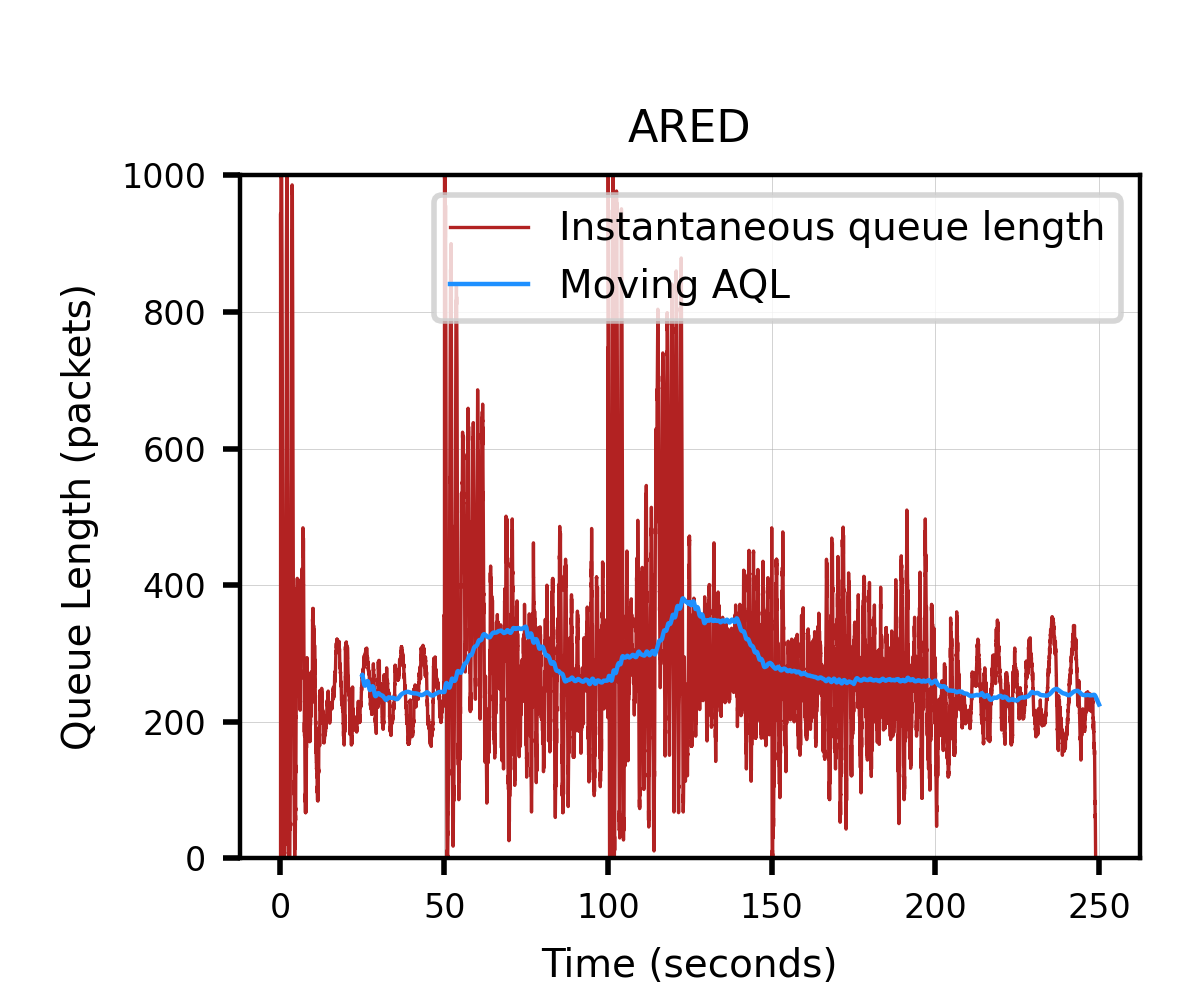}
	\includegraphics{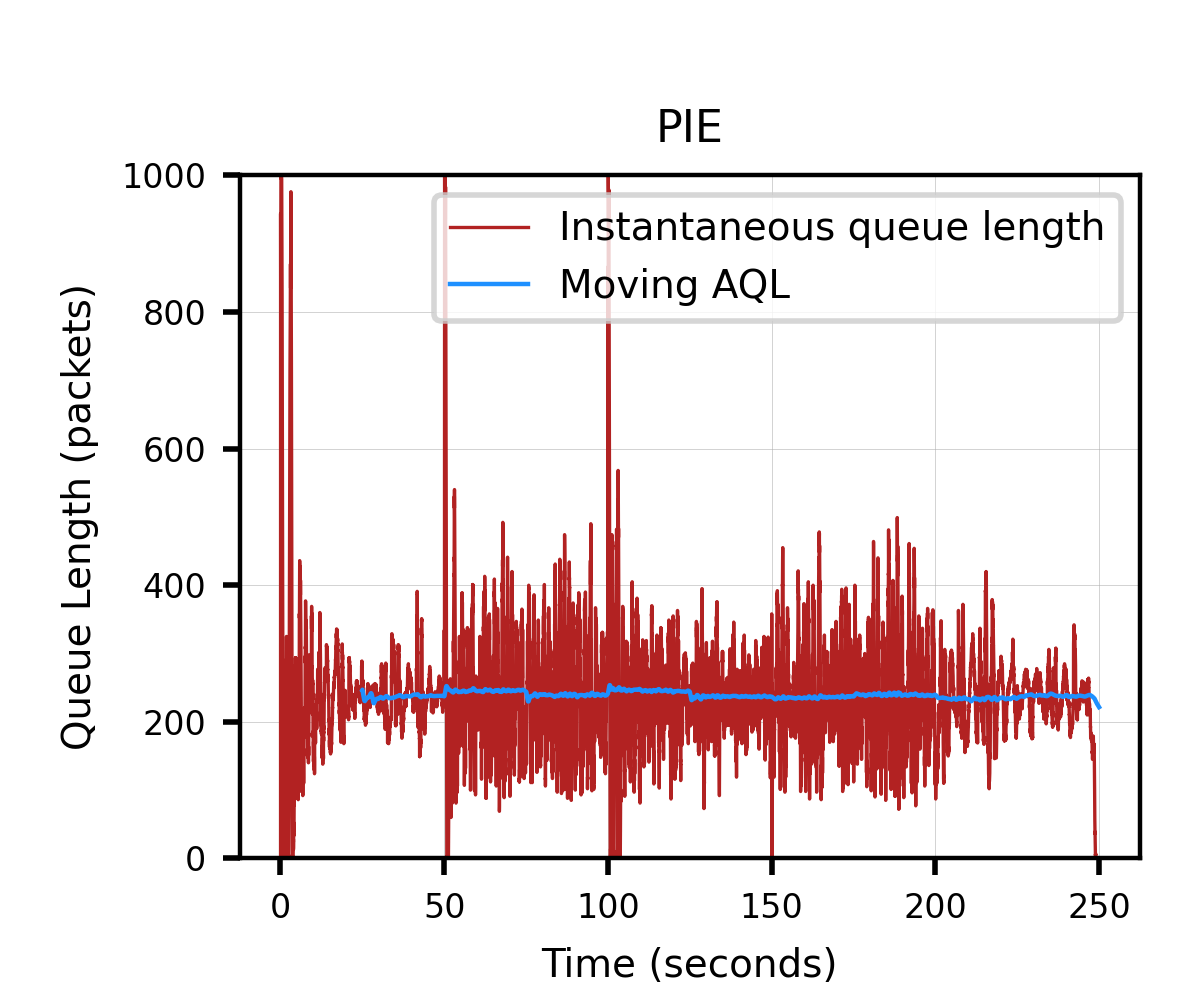}
	\includegraphics{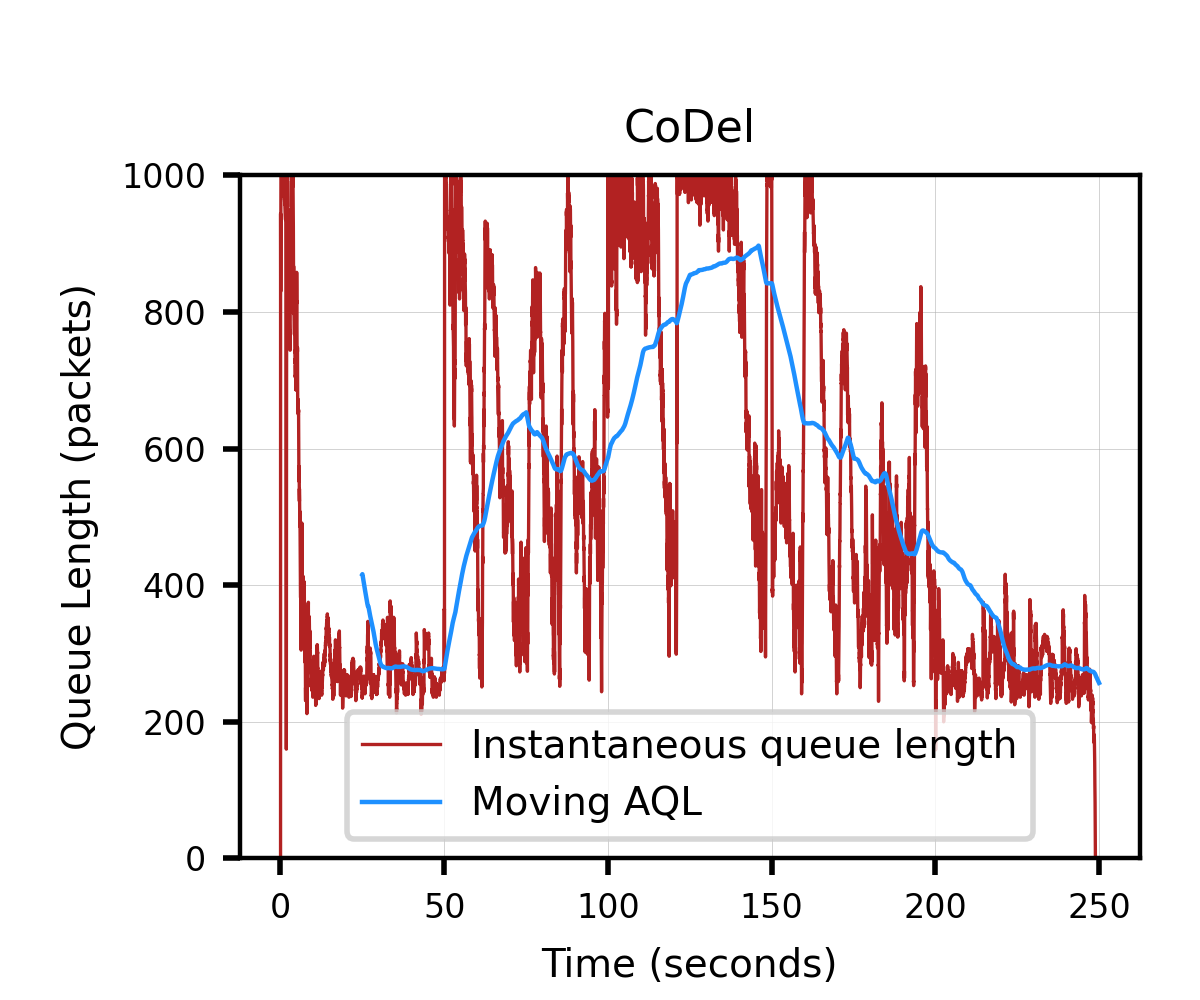}
	\includegraphics{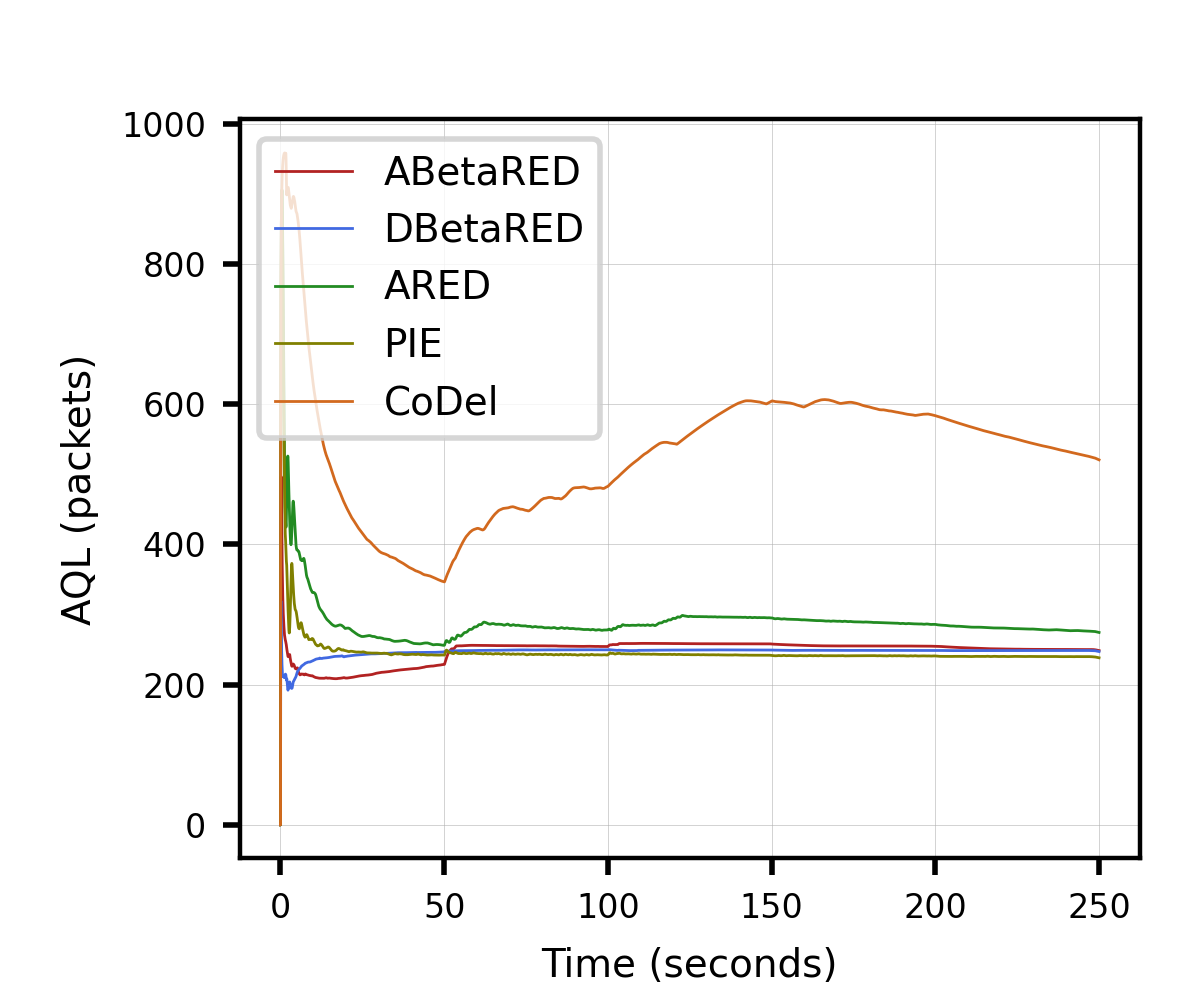}
	
	\caption{Stability comparison between related AQM algorithms when considering different levels of congestion by varying dynamically the number $N$ of active flows with $N_{\max}=800$. Other tuning parameters:  $T_{\mathrm{target}}=40\,\mathrm{ms}$ for all AQM algorithms; $\theta=0.1$ and $w=0.1$ for ABetaRED and DBetaRED. Scenario 2 on the dumbbell topology (described in Section \ref{subsec-scenarios-simulation}) is used. The bottom right pannel shows a comparison of the average queue length (for the total simulation time) of all selected AQM algorithms. The other panels illustrate the instantaneous queue length pattern, together with its moving average queue length (of the last $25$ seconds) for each of the AQM algorithms. Among all, DBetaRED exhibits the highest stability.}

	\label{fig-comparison-stability-all-aqms}
\end{figure}

As can be seen in Figure \ref{fig-comparison-all-algorithms}, when fluctuations and congestion level grow, all performance metrics of the DBetaRED algorithm outperform all other AQM algorithms. Moreover, in Figure \ref{fig-comparison-stability-all-aqms} a greater stability around the target value of the average queue length is also observed.

\section{Conclusions}\label{sec-conclusions}

We have shown that the new BetaRED algorithm is a simple, flexible and robust mechanism, which provides good stability and performance over a very wide range of parameters. However, BetaRED needs parameter tuning according to the network characteristics and congestion scenario. In order to reduce the number of user tuning parameters, the dynamic algorithms ABetaRED and DBetaRED (based on BetaRED) have been proposed and compared with benchmark algorithms such as ARED, CoDel and PIE, obtaining comparable results and even outperforming them in certain scenarios. DBetaRED stabilizes queueing length, further improves throughput and reduces packet drops, compared to other representative AQM algorithms, most notably in high-load congestion scenarios. Although the simulations carried out to obtain these conclusions have been numerous, the testing possibilities are enormous, and so there is still a lot of work to be done in this regard.

The proposed BetaRED-type algorithms are not claimed to provide a general optimal solution, since the optimality of the parameters will depend on the objective set and the characteristics of the scenarios. Therefore, one of our future research topics will be the search for optimal settings of BetaRED parameters in different concrete network scenarios, for which an in-depth analysis of mathematical models derived from the BetaRED algorithm will be necessary.

\section*{Acknowledgments}
This research work was ﬁnancially supported by FEDER/Ministerio de Ciencia e Innovación, Agencia Estatal de Investigación, grant PID2019-108654GB-I00.
A. Giménez, J.M. Amigó and J. Valero have also been partially supported by the Generalitat Valenciana (Spain), project PROMETEO/2021/063.

\bibliographystyle{ieeetr} 

\bibliography{references}

\end{document}